\documentclass[longauth]{aa}

\usepackage{amsmath,amssymb}
\usepackage{graphicx}
\usepackage{txfonts}
\usepackage{subfigure}
\usepackage{float}
\usepackage{natbib,twoopt}
\usepackage{xspace}
\usepackage{balance}
\usepackage{upgreek}
\usepackage{multirow}

\def \nat{Nature}

\def\apj{Astrophys. J.}

\def\apjl{Astrophys. J. Lett.}
\def\apjs{Astrophys. J. Suppl.}
\def\mnras{Mon. Not. R. Astron. Soc.}
\def\pasp{Publ. Astron. Soc. Pac.}
\def\pasj{Publ. Astron. Soc. Japan}
\def\physrep{Phys. Rept.}

\def\aap{Astron. Astrophys.}

\def \jcap{J. Cosm. Astro-Particle Phys.}

\def\msun{\mbox{M$_\odot$}}

\newcommand{\Mpch}{$h^{-1}\,\mbox{Mpc}$\,}
\newcommand{\hmpc}{$h^{-1}\,\mbox{Mpc}$\,}

\def\ltsim{\lower.5ex\hbox{$\; \buildrel < \over \sim \;$}}

\bibpunct{(}{)}{;}{a}{}{,} 
\DeclareGraphicsExtensions{.eps,.ps,.pdf}

\begin{document}

\title{ The VIMOS Public Extragalactic Redshift Survey (VIPERS)
  \thanks{Based on observations collected at the European Southern
    Observatory, Paranal, Chile, under programmes 182.A-0886 (LP) at
    the Very Large Telescope, and also based on observations obtained
    with MegaPrime/MegaCam, a joint project of CFHT and CEA/DAPNIA, at
    the Canada-France-Hawaii Telescope (CFHT), which is operated by
    the National Research Council (NRC) of Canada, the Institut
    National des Science de l'Univers of the Centre National de la
    Recherche Scientifique (CNRS) of France, and the University of
    Hawaii. This work is based in part on data products produced at
    TERAPIX and the Canadian Astronomy Data Centre as part of the
    Canada-France-Hawaii Telescope Legacy Survey, a collaborative
    project of NRC and CNRS. The VIPERS web site is
    http://vipers.inaf.it/.}  
Measuring non-linear galaxy bias at $z\sim0.8$}

\author{
C.~Di Porto\inst{\ref{9}}
\and E.~Branchini\inst{\ref{10},\ref{28},\ref{29}}
\and J.~Bel\inst{\ref{2}}
\and F.~Marulli\inst{\ref{17},\ref{9},\ref{18}}
\and M.~Bolzonella\inst{\ref{9}}
\and O.~Cucciati\inst{\ref{9}}
\and S.~de la Torre\inst{\ref{4}}
\and B.~R.~Granett\inst{\ref{2}}
\and L.~Guzzo\inst{\ref{2},\ref{27}}
\and C.~Marinoni\inst{\ref{7}}
\and L.~Moscardini\inst{\ref{17},\ref{9},\ref{18}}
\and U.~Abbas\inst{\ref{5}}
\and C.~Adami\inst{\ref{4}}
\and S.~Arnouts\inst{\ref{4},\ref{6}}
\and D.~Bottini\inst{\ref{3}}
\and A.~Cappi\inst{\ref{9},\ref{30}}
\and J.~Coupon\inst{\ref{12}}
\and I.~Davidzon\inst{\ref{17},\ref{9}}
\and G.~De Lucia\inst{\ref{13}}
\and A.~Fritz\inst{\ref{3}}
\and P.~Franzetti\inst{\ref{3}}
\and M.~Fumana\inst{\ref{3}}
\and B.~Garilli\inst{\ref{4},\ref{3}}
\and O.~Ilbert\inst{\ref{4}}
\and A.~Iovino\inst{\ref{2}}
\and J.~Krywult\inst{\ref{15}}
\and V.~Le Brun\inst{\ref{4}}
\and O.~Le F\`evre\inst{\ref{4}}
\and D.~Maccagni\inst{\ref{3}}
\and K.~Ma{\l}ek\inst{\ref{16}}
\and H.~J.~McCracken\inst{\ref{19}}
\and L.~Paioro\inst{\ref{3}}
\and M.~Polletta\inst{\ref{3}}
\and A.~Pollo\inst{\ref{22},\ref{23}}
\and M.~Scodeggio\inst{\ref{3}}
\and L.~A.~M.~Tasca\inst{\ref{4}}
\and R.~Tojeiro\inst{\ref{11}}
\and D.~Vergani\inst{\ref{25}}
\and A.~Zanichelli\inst{\ref{26}}
\and A.~Burden\inst{\ref{11}}
\and A.~Marchetti\inst{\ref{2},\ref{1}}
\and D.~Martizzi\inst{\ref{40}}
\and Y.~Mellier\inst{\ref{19}}
\and R.~C.~Nichol\inst{\ref{11}}
\and J.~A.~Peacock\inst{\ref{14}}
\and W.~J.~Percival\inst{\ref{11}}
\and M.~Viel\inst{\ref{13},\ref{41}}
\and M.~Wolk\inst{\ref{19}}
\and G.~Zamorani\inst{\ref{9}}
}
\institute{
 INAF - Osservatorio Astronomico di Bologna, via Ranzani 1, I-40127 Bologna, Italy \label{9}
\and Dipartimento di Matematica e Fisica, Universit\`{a} degli Studi Roma Tre, via della Vasca Navale 84, I-00146 Roma, Italy \label{10}
\and INFN - Sezione di Roma Tre, via della Vasca Navale 84, I-00146 Roma, Italy \label{28}
\and INAF - Osservatorio Astronomico di Roma, via Frascati 33, I-00040 Monte Porzio Catone (RM), Italy \label{29}
\and INAF - Osservatorio Astronomico di Brera, Via Brera 28, 20122 Milano, via E. Bianchi 46, I-23807 Merate, Italy \label{2}
\and Dipartimento di Fisica e Astronomia - Universit\`{a} di Bologna, viale Berti Pichat 6/2, I-40127 Bologna, Italy \label{17}
\and INFN - Sezione di Bologna, viale Berti Pichat 6/2, I-40127 Bologna, Italy \label{18}
\and Aix Marseille Universit\'e, CNRS, LAM (Laboratoire d'Astrophysique de Marseille) UMR 7326, 13388, Marseille, France  \label{4}
\and INAF - Osservatorio Astronomico di Trieste, via G. B. Tiepolo 11, I-34143 Trieste, Italy \label{13}
\and  INFN - Istituto Nazionale di Fisica Nucleare, Via Valerio 2, I-34127 Trieste, Italy \label{41}
\and Dipartimento di Fisica, Universit\`a di Milano-Bicocca, P.zza della Scienza 3, I-20126 Milano, Italy \label{27}
\and Centre de Physique Th\'eorique, UMR 6207 CNRS-Universit\'e de Provence, Case 907, F-13288 Marseille, France \label{7}
\and Astronomical Observatory of the Jagiellonian University, Orla 171, 30-001 Cracow, Poland \label{22}
\and National Centre for Nuclear Research, ul. Hoza 69, 00-681 Warszawa, Poland \label{23}
\and INAF - Osservatorio Astronomico di Torino, I-10025 Pino Torinese, Italy \label{5}
\and Canada-France-Hawaii Telescope, 65--1238 Mamalahoa Highway, Kamuela, HI 96743, USA \label{6}
\and INAF - Istituto di Astrofisica Spaziale e Fisica Cosmica Milano, via Bassini 15, I-20133 Milano, Italy\label{3}
\and Laboratoire Lagrange, UMR7293, Universit\'e de Nice Sophia-Antipolis,  CNRS, Observatoire de la C\^ote d'Azur, 06300 Nice, France \label{30}
\and Institute of Astronomy and Astrophysics, Academia Sinica, P.O. Box 23-141, Taipei 10617, Taiwan\label{12}
\and Institute of Physics, Jan Kochanowski University, ul. Swietokrzyska 15, 25-406 Kielce, Poland \label{15}
\and Department of Particle and Astrophysical Science, Nagoya University, Furo-cho, Chikusa-ku, 464-8602 Nagoya, Japan \label{16}
\and Institute d'Astrophysique de Paris, UMR7095 CNRS, Universit\'{e} Pierre et Marie Curie, 98 bis Boulevard Arago, 75014 Paris, France \label{19}
\and Universit\"{a}tssternwarte M\"{u}nchen, Ludwig-Maximillians Universit\"{a}t, Scheinerstr. 1, D-81679 M\"{u}nchen, Germany \label{24}
\and Max-Planck-Institut f\"{u}r Extraterrestrische Physik, D-84571 Garching b. M\"{u}nchen, Germany \label{20}
\and Institute of Cosmology and Gravitation, Dennis Sciama Building, University of Portsmouth, Burnaby Road, Portsmouth, PO1 3FX, UK \label{11}
\and INAF - Istituto di Astrofisica Spaziale e Fisica Cosmica Bologna, via Gobetti 101, I-40129 Bologna, Italy \label{25}
\and INAF - Istituto di Radioastronomia, via Gobetti 101, I-40129 Bologna, Italy \label{26}
\and SUPA - Institute for Astronomy, University of Edinburgh, Royal Observatory, Blackford Hill, Edinburgh, EH9 3HJ, UK \label{14}
\and Universit\`{a} degli Studi di Milano, via G. Celoria 16, I-20130 Milano, Italy \label{1}
\and  Department of Astronomy, University of California, Berkeley, CA 94720 \label{40}
}

\date{Received --; accepted --}

\abstract 
{}
{We use the first release of the VImos Public Extragalactic Redshift Survey 
of galaxies (VIPERS) of $ \sim 50,000$ objects to measure 
the biasing relation between galaxies and mass in the redshift range $z=[0.5,1.1]$.} 
{We estimate the 1-point distribution function [PDF] of  VIPERS galaxies from 
 counts in cells and, assuming a model for the mass PDF, we 
 infer their mean bias relation. 
 The reconstruction of the bias relation  is performed through a novel method that 
 accounts for Poisson noise, redshift distortions, inhomogeneous sky coverage. and other selection effects.
 With this  procedure we constrain galaxy bias and its deviations from linearity
down to scales as small as 4 \Mpch and out to $ z=1.1 $.
}
{We detect small (up to $ 2$ \%) but statistically significant (up to 3 $\sigma$) deviations from linear bias. 
The mean biasing function is close to linear in regions above the mean density. 
The mean slope of the biasing relation is a proxy to the linear bias parameter. 
This slope increases  with luminosity, which is in agreement with results of previous analyses.
We detect a strong bias evolution only for $z > 0.9$, which is in agreement with some, but not all, 
previous studies.
We also detect a significant increase of the bias with the scale, from 4 to 8 \hmpc, now seen for the first time out to $z=1$.
The amplitude of non-linearity depends on redshift, luminosity,  and scale, but no
clear trend is detected. 
Owing to the large cosmic volume probed by VIPERS, we find that the
mismatch between the previous estimates of bias at $z\sim 1$ from zCOSMOS and VVDS-Deep 
galaxy samples is fully accounted for by cosmic variance.
}
{
The results of our work confirm the importance of going beyond 
the over-simplistic linear bias hypothesis showing that non-linearities can be accurately measured through the applications 
of the appropriate  statistical tools to existing datasets like VIPERS.
}

\maketitle

\keywords{Cosmology: observations -- Cosmology: large-scale
      structure of Universe -- Surveys -- Galaxies: evolution }
    
\authorrunning{C. Di Porto et al.}
\titlerunning{The bias of  VIPERS galaxies}


\section{Introduction}
\label{sec:intro}

Galaxies do not perfectly trace mass. The long known proof 
is that galaxy clustering depends on
properties of galaxies such as luminosity, colour, morphology,
stellar mass, and so on (e.g. \citealt{szapudi00,hawkins01,norberg01, norberg02,zehavi02, zehavi011,meneux09,marulli13}) and not solely on the underlying mass distribution.
Differences in clustering properties are caused by the physical  processes that regulate 
the formation and evolution of galaxies and should disappear when averaging over scales
much larger than those affected by  these processes.

Modelling the physics of galaxy formation, or at least its impact on the
{\it bias} relation, 
is of paramount importance to extract cosmological information from the spatial distribution
of galaxies. 
Indeed, the large-scale structure of the Universe as traced by  galaxies  
is one of the most powerful cosmological probes as
 testified by the increasing number of large galaxy redshift surveys either ongoing, such as Boss \citep{anderson12}, 
 DES\footnote{www.darkenergysurvey.com}, and VIPERS \citep{guzzo13} or those planned for the near future, such as eBOSS\footnote{ http://www.sdss3.org/future/}, DESI \citep{schlegel11}, and Euclid \citep{euclidRB}\footnote{http://www.euclid-ec.org/}.
These surveys are designed to address several important questions both in cosmology and in galaxy evolution theory. Chief among them is the origin of  the 
accelerated expansion of the Universe.

It has recently been realised that geometry tests based on standard candles and standard rulers
can trace the expansion history of the Universe but cannot identify the cause of the accelerated expansion,
which
can be obtained either by advocating a dark energy component or by modifying 
the gravity theory (e.g. \citealt{wang08}). To break this degeneracy one needs independent observational tests. 
These are provided by  the build-up of structures over cosmic time \citep{guzzo08}.
The analysis of large-scale structures in galaxy distribution allows us to perform these two tests at one time. The baryonic acoustic oscillation peaks in the two point statistics provide a standard ruler to perform geometry test
(e.g. \citealt{seo03,percival07,cabre09,reid012})  whereas the apparent radial distortions in galaxy clustering 
caused by peculiar motions that are gravitationally induced  allow us to measure the rate at which cosmic structures grow.
Since both tests rely on baryonic structures, the knowledge of the bias relation is mandatory to probe the 
underlying mass distribution and set cosmological constraints.
Notwithstanding, a clustering statistics  that is in principle bias insensitive has  been recently proposed by
\cite{bel14} and applied to VIPERS data \citep{bel14b}.

Galaxy bias is not just a nuisance parameter in the quest for the world model.
This bias also represents an opportunity to constrain models of galaxy evolution as it encodes important information
about the physical processes that regulate the evolution of stars and galaxies.
Therefore, it is important to model galaxy bias by establishing its link  to the relevant astrophysical processes that regulate galaxy evolutions.In a recent review, \cite{baugh13} has classified galaxy evolution models into two categories. 
The so-called { empirical models} belong to the first category. These authors use theoretically motivated relations to model galaxy distribution 
from halos extracted from $N$-body simulations. The two most popular 
schemes to populate halos with galaxies are halo occupation distribution (HOD; e.g. \citealt{cs02,zheng05}) and sub-halo abundance matching (SHAM; e.g. \citealt{vo04,conroy06}). 
The second category is represented by  { physical models} in which 
the  processes that regulate the evolution of baryons are explicitly considered to
link them to the host dark matter structures. This approach is at the heart of the semi-analytic models of galaxy formation 
(e.g. \citealt{wf91,bower06,db07}).
In most cases these models have been used to estimate  galaxy bias from clustering statistics
such as galaxy counts or 2-point correlation functions.  The results indicate that the accuracy in both types of models
is one of the main limitations in constraining dark energy or modified gravity from current and, even more so, future observational campaigns
\citep{contreras013}. 

Alternatively, one can adopt a purely phenomenological approach and use an
operational definition of the bias in terms of  map between 
the density fluctuations of mass, $\delta$ and  galaxies, $\delta_g$ smoothed on the same scale.
This approach assumes that galaxy bias is a local process that depends on the local mass density only.
Many studies further assume that the bias relation is 
linear and deterministic, so that galaxy bias can be quantified by a single {\it linear bias} parameter $b$: $\delta_g=b \delta$.
The concept of linear bias has played an important role in cosmology and many results have been obtained using this
assumption, which is known to be unphysical as it allows negative densities. Also, this assumption
has no
justification at the relatively small scales of interest to the study of galaxy formation processes, which depend on many physical parameters and
on large scales due to the presence of neutrinos \citep{villaescusa14}.
In fact, the bias is constant only on scales larger than about 40 \hmpc \citep{MG11}.
Indeed, galaxy bias can be more conveniently
described within a probabilistic framework as proposed by \cite{dl99} and recently reformulated in the context of the halo model \citep{cacciato012}.

From the phenomenological viewpoint,  bias has been extensively investigated from counts in cells statistics, 
weak gravitational lensing, and galaxy clustering. The latter is 
probably most popular approach. It is typically based on
2-point statistics and on the assumption of linear bias
\citep{norberg01, norberg02, zehavi05,coil06,bp07,nuza012,alhambra13,skibba13,marulli13}. 
A comparatively smaller number of studies searched for deviations from the linear and deterministic bias
either using 2-point \citep{tegmark99}
or higher order statistics \citep{verde02,gaztanaga05,kayo04,nishimichi07, swanson08}.

Gravitational lensing in the weak field regime has also been exploited to constrain galaxy bias.
{ In particular, within the limit of scale-independent bias on large scales, 
weak lensing and galaxy clustering can be combined to estimate the linear bias parameter in a manner which is
independent of the amplitude of density fluctuations \citep{amara12,pujol16,chang16}.} On smaller scales
weak lensing was also used to measure the scale dependence of galaxy bias 
\citep{hoekstra02, simon07, jullo12,comparat13}, although this effect is degenerate 
with bias stochasticity, i.e. the fact that galaxy bias might not be solely determined by 
the local mass density.

The most natural way to study a possible scale dependence (or non-linearity) of galaxy bias is 
in a probabilistic framework  by means of 
counts in cells statistics  \citep{Sigad00} since in this case one can separate deviations 
from linear bias and the presence of an intrinsic scatter in the bias relation.
This approach was used to estimate the bias of galaxies in the  PSC$z$  \citep{branchini01}, VVDS 
(\citealt{marinoni05}, hereafter M05), and zCOSMOS (\citealt{kovac011}, hereafter K11) catalogues as well as  
the relative bias of blue versus red galaxies in the  2 degrees field galaxy redshift survey (2dFGRS; 
 \citealt{colless01,Wild}).
Despite some disagreement, results obtained at low redshift ($z<0.5$) generally indicate 
that, at least for some types of galaxies, the bias is stochastic, scale dependent 
and, therefore, non-linear. However.
The situation at $z>0.5$ is less clear. Gravitational lensing studies either focused on very bright objects to probe the baryonic acoustic oscillations \citep{comparat13}
or on galaxies in the COSMOS field \citep{jullo12}; these studies found  no evidence for stochasticity but, in the case of \cite{jullo12}, detected a 
significant scale dependence of galaxy bias.
This conflicting evidence shows a lack of accuracy in current estimators for galaxy bias
that is a serious warning for precision cosmology. { This is especially true considering that
this is the range that will be  probed by next generation surveys
that have the potential to trace both the redshift and scale dependence of galaxy bias 
\citep{dp12a,dp12b}}

 The results obtained so far that focus on counts in cells
provide some conflicting evidence. 
In M05 authors analysed galaxies in the VVDS-Deep catalogue
over an area of $0.4 \times 0.4$ deg and
found significant deviations from linearity.
The estimated effective linear bias parameter showed little evolution with redshift. In contrast,
the biasing relation of zCOSMOS galaxies measured by K11 over a region of about 1.52 deg$^2$
turned out to be close to linear and rapidly evolving with the redshift.
The tension between these results is paralleled  by the observed differences in the spatial correlation properties of the two samples,
with the 2-point correlation function in zCOSMOS systematically higher than that of VVDS galaxies
(see e.g. \citealt{meneux09}). 
Owing to the large cosmic variance in the two samples, a rather small galaxy sample was proposed as the source
of this mismatch, so a larger galaxy sample should be used to settle the issue.

The Vimos Public Extragalactic Redshift Survey [VIPERS] \citep{guzzo13}
has a depth similar to the zCOSMOS survey
but with a much larger area of  24 deg$^2$. Its volume is comparable to that of 2dFGRS  and 
is large enough to significantly reduce the impact of the cosmic variance (see Appendix in \citealt{fritz14}).
We adopt the same approach as M05 and K11 and estimate galaxy bias from counts in cells.
To do so we use a novel estimator that accounts for the effect of discrete sampling, allowing us to 
use small cells and probe unprecedented small scales that are more affected by the physics of galaxy formation.

The layout of the paper is as follows. In Section \ref{sec:data} we describe both the real and mock datasets
used in this work. In Section \ref{sec:theory} we introduce the formalism used to characterise galaxy bias 
and the estimators used to measure this bias from a galaxy redshift survey. In Section \ref{sec:esources} we assess the 
validity of the estimator and use mock galaxy catalogues to gauge random and systematic errors.
We present our results in Section  \ref{sec:results} and compare these with those of other analyses
in Section~\ref{sec:comparison}. The main conclusions are drawn in Section \ref{sec:conclusions}

Throughout this paper we assume a flat $\Lambda$CDM
universe  ($\Omega_m$, $\Omega_{\Lambda}$, $\sigma_8$)= (0.25; 0.75; 0.9). 
Galaxy magnitudes are given in the AB system and, unless otherwise stated, 
computed assuming   $h\equiv H_0/100\, {\rm km\, s^{-1} Mpc^{-1} }=1 $.The high value of $\sigma_8$  has little impact on our analysis since our results 
can be rescaled to different values of $\sigma_8$ that are more consistent with current cosmological constraints.
The dependence of the magnitude upon $h$ is expressed as 
$M=M_h-5 \log(h)$, where $M_h$ is the absolute magnitude computed for a given $h$ value.


\section{Datasets}
\label{sec:data}

The results in this paper are based on the first release of the 
VIPERS galaxy catalogue \citep{garilli13}. 
Random and systematic errors
were computed using a set of 
simulated galaxy catalogues  mimicking the real catalogue and its observational selections.
Both, the real and mock samples are described in this Section.

\subsection{Real data}
\label{sec:realdata}
The VIMOS Public Extragalactic Redshift Survey  is
an ongoing ESA Large Programme aimed at measuring spectroscopic redshifts for
about $10^5$  galaxies at redshift $0.5 < z < 1.2$ and beyond.
The
galaxy target sample is selected from the `T0005' release of the  Canada-France-Hawaii
Telescope Legacy SurveyWide (CFHTLS-Wide)
 optical photometric
catalogue\footnote{http://terapix.iap.fr/cplt/oldSite/Descart/CFHTLS-T0005-Release.pdf}. 
VIPERS covers 24 deg$^2$ on
the sky, divided over two areas within the W1 and W4 CFHTLS
fields. Galaxies are selected to a limit of 
$I_{AB} < 22.5$, further applying
a simple and robust colour preselection to efficiently
remove galaxies at $z < 0.5$. This colour cut and the 
adopted observing strategy \citep{scodeggio09}
allow us to double the galaxy
sampling rate  with respect to a pure magnitude-limited sample. 
At the same time, the area and depth of
the survey result in a relatively large volume, $5 \times 10^7 \  \ h^{-3}$ Mpc$^3$,
which is analogous to that of the 2dFGRS 
at $z\sim0.1$. 
VIPERS spectra are collected with the VIMOS
multi-object spectrograph \citep{lefevre03} at moderate resolution
(R = 210) using the LR Red grism, providing a wavelength
coverage of 5500-9500 $\AA$
and a typical radial velocity error
of ${\sigma_v}=141 (1+z)$ km s$^{-1}$.

The full VIPERS area of 24 deg$^2$ is covered through a mosaic of 288 VIMOS pointings.
A complete description of the
survey construction, from the definition of the target sample to the
actual spectra and redshift measurements, is given in \cite{guzzo13}.
The dataset used in this and other papers of the early
science release represent the VIPERS Public Data Release 1
(PDR-1) catalogue that includes 55359 redshifts (27935 in W1 and 27424
in W4), i.e.  64\% of the final survey in terms of covered area \citep{garilli13}. A quality
flag was assigned to each object in the process of determining
their redshift from the spectrum, which quantifies the reliability of
the measured redshifts. In this analysis, we use only galaxies with
flags 2 to 9.5, which corresponds to a sample with a redshift confirmation
rate of 90\%. 

Several observational effects need to be taken into account to investigate the 
spatial properties of the underlying population of galaxies.

{\it i)} Selection effects along the radial direction
are driven by the flux limit  nature of the survey and, at $z<0.6$, by the colour preselection strategy.
We use volume-limited (luminosity-complete) galaxy subsamples that we obtain 
by selecting galaxies brighter than a given magnitude threshold in a given redshift interval. 
We adopted a redshift-dependent luminosity cut  of the form $M_B(z) = M_0 - z$ that should account for 
the luminosity evolution of galaxies (e.g. \citealt{zucca09}). The value of the threshold is 
set to guarantee that the selected sample is $>90$ \% complete 
within the given redshift interval. In this sense each subsample is volume limited and luminosity complete.
This $z$-dependent luminosity cut is very popular and has been adopted in other papers (see e.g. K11). However, other works used
different types of cuts, either ignoring any dependence on redshift  (such as in M05; \citealt{coil08}) 
 or assuming a different functional form for the redshift evolution (e.g. \citealt{alhambra13}).
Adopting an incorrect  luminosity evolution would generate a spurious
radial gradient in the mean density of the objects and a wrong $z-$dependence in the 
galaxy bias. To minimise the impact of this potential bias, we carry out our analysis in relatively 
narrow redshift bins, so that adopting any of the aforementioned luminosity cuts would produce similar results,
as we verified. The robustness of our result to the choice of the magnitude cut can be { tested {\it a posteriori}}. 
Figure~\ref{fig:bias_compf} shows that the difference between estimates obtained with a $z$-dependent cut (filled red dot)
 and with a $z$-independent cut (open red dot) are smaller than the total random errors.

Selection effects induced by the colour preselection strategy were determined from the comparison 
between the spectroscopic and photometric samples \citep{guzzo13, delatorre13,fritz14} 
and are accounted for by assigning to each galaxy an appropriate statistical weight dubbed colour sampling rate (CSR).

{\it ii)} The surveyed area presents regular gaps due to the specific
footprint of the VIMOS spectrograph that creates  a  pattern of 
rectangular regions, called pointings, separated by
gaps where no spectra are taken. Superimposed on this pattern are 
unobserved areas resulting from bright stars and technical and mechanical 
problems during observations. We discuss our strategy
to take into account this effect in our  counts in cells analysis  in the following (see \citealt{cucciati14}, for a more detailed study).

{\it iii)} In each pointing, slits are assigned to a number of potential targets that meet the survey
selection criteria \citep{bottini05}. Given the surface density of the targeted population, the multiplex capability
of VIMOS, and the survey strategy, a fraction of about 45\% of
the parent photometric sample can be assigned to slits. We define the fraction of targets that have a measured 
spectrum as the target sampling rate (TSR) and the fraction of observed
spectra with reliable redshift measurement as the spectroscopic
sampling sate (SSR). Both functions are roughly independent of galaxy magnitude except the SSR, which decreases for $I_{AB} > 21.0$, as shown in 
Fig. 12 of \cite{guzzo13}.


All these selection effects are thoroughly discussed and quantitatively assessed  by \cite{delatorre13}.
We make no attempt to explicitly correct for these effects individually. Instead, we assess their impact on the estimate of galaxy bias in Section~\ref{sec:esources} using the mock galaxy catalogues described below.

For the scope of our analysis, the main advantages of VIPERS are the relatively dense sampling of tracers, which
allows us to probe density fluctuations down to scales comparable to those affected by galaxy evolution processes, 
and the large volume that, as discussed in the previous Section,
allows us to reduce the impact of cosmic variance considerably  with respect to previous
estimates of galaxy bias at $z\sim1$.

The parent PDR-1 VIPERS sample contains 45871 galaxies with reliable redshift measurements. 
Here we restrict our analysis in the redshift range $z=[0.5,1.1]$ since the number density of objects at larger distances is too
small to permit a robust estimate of galaxy bias.
To investigate the possible dependence of galaxy bias on luminosity and redshift, we partitioned 
the catalogue into subsamples by applying a series of cuts in both magnitude and redshift.

\begin{figure}
\includegraphics[width=0.55\textwidth]{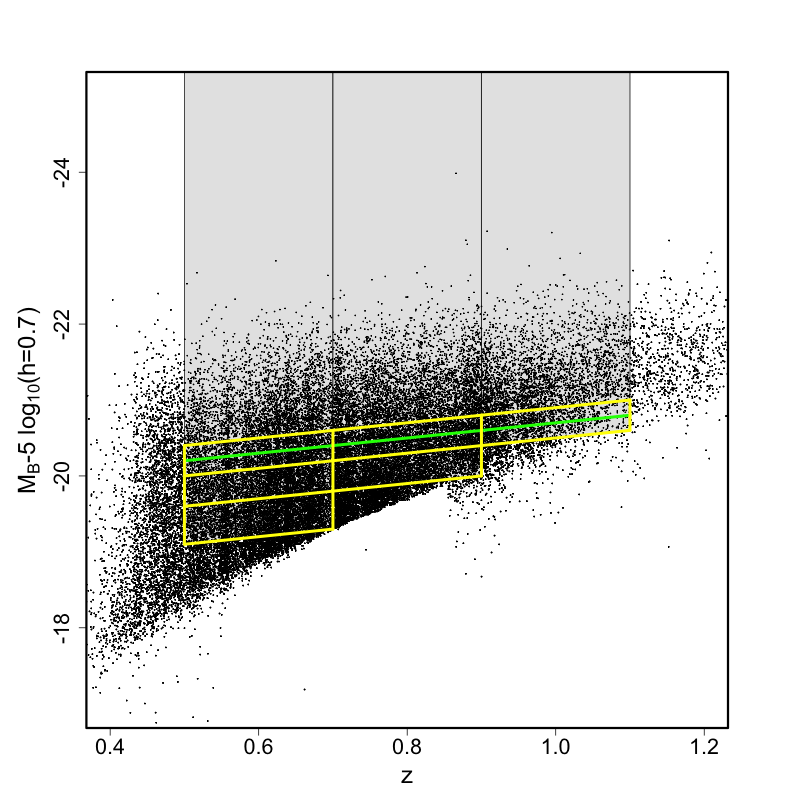}
\caption{Luminosity selection as a function of redshift. The black dots show the W1 and W4 VIPERS galaxies (with spectroscopic redshift flag between 2 and 9.5). Yellow lines represent the principal magnitude cuts applied in every redshift bin. The green line represents the cut $M_0=-19.7-z$ made to compare our results to those  of K11.
}
\label{fig:MBvsz}
\end{figure}

The complete list of subsamples considered in this work is presented in Table~\ref{tab:cats}.
We considered three redshift bins ($z=[0.5,0.7], \, [0.7,0.9], \, [0.9,1.1]$) and applied 
different luminosity cuts that we obtained by compromising between the 
need of maximising both completeness and number of objects. 
Different luminosity cuts within each redshift bin allow us to study the luminosity dependence of galaxy bias at different redshifts. 
The magnitude cuts, M$_{\rm B}=-19.5-z-5\log(h)$ and $-19.9-z-5\log(h),$ that run across the whole redshift range are used to investigate a possible evolution of galaxy bias.
In the Table the subsamples are listed in groups. The first three groups indicate subsamples in the three redshift bins. 
The last group indicates 
subsamples that are designed to match the luminosity cuts performed by K11
(M$_{\rm B} = -20.5 -z-5\log(h=0.7)=-19.72 -z-5\log(h)$) and by M05 (M$_{\rm B} = -20.0 -5\log(h)$.
The most conservative cut  M$_{\rm B}=-19.5-z-5\log(h)$ guarantees 90 \% completeness out 
to $z=1$ for the whole galaxy sample and higher for late type objects (see Fig. \ref{fig:MBvsz}). 

Since the analysis presented in this work is based on cell count statistics, a useful figure of merit is represented 
by the number of independent spheres that can be accommodated within the volume of the survey.
Considering intermediate cells with a radius of 6 \hmpc, the number of such independent cells is
$N= 3869, \, 5527, \, 6964$ in the three redshift intervals $z=[0.5,0.7], \, [0.7,0.9], \, [0.9,1.1]$, respectively.

 \begin{table}
\caption{VIPERS subsamples. }     
\label{tab:cats}      
\centering          
\begin{tabular}{c c c c}     
\hline\hline    
$z$-range & M$_{\rm B}$- cut & $n_{\rm VIPERS}$ & $n_{\rm mock}$ \\
 & M$_{\rm B}-5\log(h)$ & $10^{-3} h^3 \, {\rm Mpc}^{-3}$ & $10^{-3} h^3 \, {\rm Mpc}^{-3}$   \\
\hline
0.5 - 0.7 & $-18.6-z$ &  4.78 & 4.36  \\
0.5 - 0.7 & $-19.1-z$ &  3.16 & 2.43  \\
0.5 - 0.7 & $-19.5-z$ &  2.10 & 1.37  \\
0.5 - 0.7 & $-19.9-z$ &  1.24 & 0.68  \\
\hline
0.7 - 0.9  & $-19.1-z$ & 2.71 & 2.55  \\
0.7 - 0.9  & $-19.5-z$ & 1.86 & 1.47  \\
0.7 - 0.9  & $-19.9-z$ & 1.07 & 0.72  \\
\hline
0.9 - 1.1  & $-19.5-z$ & 0.62 & 0.63  \\
0.9 - 1.1  & $-19.9-z$ & 0.42 & 0.43  \\
\hline
0.5 - 0.7  & $-19.7-z$ & 1.64 & 1.36  \\
0.7 - 0.9  & $-19.7-z$ & 1.13 & 1.05 \\
0.9 - 1.1  & $-19.7-z$ & 0.53 & 0.53  \\
0.7 - 0.9  & $-20.0$    & 1.42 & 1.49  \\
\hline \hline
\end{tabular}
\tablefoot{ Col. 1: redshift range. Col. 2: B-band magnitude cut (computed for $h=1$). 
Col  3: galaxy number density in the real VIPERS sub-catalogues.  
Col  4: galaxy number density in the HOD-mock VIPERS sub-catalogues. 
In the {\it Parent} mock catalogue the number density is a factor 
$\sim 3.7$ larger. 
 Cells fully contained in the surveyed volume (i.e. not overlapping with gaps or 
empty areas) contain $\sim 40$ \%  more objects on average.}

\end{table}



\subsection{Mock datasets}
\label{sec:mockdata}

We 
 considered a suite of  mock galaxy catalogues mimicking the 
real PDR-1 VIPERS catalogue to assess our ability to measure the mean biasing 
function and evaluate random and systematic errors.

We used  two different types of mock galaxy catalogues.
We based the 
bulk of our error analysis on the first mock galaxy catalogue, which is described in  detail in \cite{delatorre13}.
In this  set of mocks, synthetic galaxies are obtained by applying the  HOD technique
to the dark matter halos extracted from  the MultiDark N-body simulation
\citep{prada012}
of a flat $\Lambda$CDM
universe  with ($\Omega_m$, $\Omega_{\Lambda}$, $\Omega_b$,
$h$, $n$, $\sigma_8$)= (0.27; 0.73; 0.0469; 0.7; 0.95; 0.82). 
Since the resolution of the parent simulation was too poor to simulate galaxies in the magnitude range 
sampled by VIPERS,  \cite{dltp012} applied an original technique to resample the halo field
to generate sub-resolution halos down to a mass of  M $= 10^{10} h^{-1}$ \msun. These halos 
were HOD populated with mock galaxies by tuning the free parameters to match the 
spatial 2-point correlation function of VIPERS galaxies  \citep{delatorre13}.
Once populated with HOD galaxies, the various outputs were rearranged to obtain
26 and 31 independent light cones mimicking the W1 and W4 fields of VIPERS
and their geometry, respectively. In our analysis we considered 26 W1+W4 mock samples. 
They constitute our set of {\it Parent} mock catalogues, as opposed to the 
{\it Realistic} 
mock catalogues that we obtain from the 
{\it Parent} set by applying the various selection effects (VIPERS footprint mask besides TSR, SSR, and CSR)
and by adding Gaussian errors to the redshifts to mimic the random error in the measured spectroscopic redshifts. 
The  mock catalogues were built assuming a constant SSR whereas, as we pointed out, this is a 
declining function of the apparent magnitude. However, the dependence is weak and 
only affects faint objects, i.e. preferentially objects at large redshifts. For this reason we decided 
to explicitly include this dependence by selectively removing objects, starting from the faintest and
moving towards brighter objects until we match the observed  SSR(m) \citep{guzzo13}.

The average galaxy number densities in the mocks are listed in Column 4 of 
Table~\ref{tab:cats}. For $z \le 0.9$ the number density in the mocks is similar 
or somewhat smaller than in the real catalogue. This discrepancy increases with the luminosity and 
probably originates from the uncertainty in the procedure to HOD-populate halos with bright mock galaxies.
The consequence for our analysis is an overestimation of the random errors in the measurement of  the bias of 
VIPERS galaxies. 
At higher redshift the trend is reversed; the number density of objects in the mocks is 
systematically larger than in the real catalogue. In this case, to avoid underestimating errors, we randomly diluted the galaxies in the mocks. Hence the perfect match of number densities in the redshift bin 
$z=[0.9,1.1]$, as shown in the table.

On the smallest scale investigated in this paper, $R=4$ \hmpc, the 
second-order statistics of simulated galaxies and 
the variance of the galaxy density field are underestimated 
by $\sim 10$ \% \citep{bel14b}. 
Therefore, 
to check the robustness of our bias estimate to the galaxy model used to generate the mock catalogues 
and to the underlying cosmological model, we  considered  a second set of mocks.
These were obtained from the Millennium $N$-body simulation \citep{springel05}
 of a flat $\Lambda$CDM
universe  with ($\Omega_m$, $\Omega_{\Lambda}$, $\Omega_b$,
$h$, $n$, $\sigma_8$)= (0.25; 0.75; 0.045; 0.73; 1.00; 0.9)
and using  the  semi-analytic technique of \cite{db07}, an alternative to the HOD.
As a result of the limited size of the computational box, it was possible to create
light cones with an angular size of 7 $\times$  1 deg$^2$, i.e. smaller than the 
individual W1 and W4 fields. Overall, we considered 
$26+26$ reduced versions of the W1+W4 fields.
From these light cones we created a corresponding number of {\it Realistic} mock catalogues.

{ Robustness tests that involve both types of mock catalogues were restricted to a limited number of 
samples (one for each redshift bin). In these tests we simply compared the errors in the bias 
estimates after accounting for the larger cosmic variance in the Millennium mocks.   
Since these robustness tests turned out to be successful in the sense that errors estimated with 
the two sets of mocks turned out to be consistent with each other,  
we do not mention these mocks again and, for the rest of the paper,
fully rely on the error estimates obtained from the HOD mocks.}


\section{Theoretical background}
\label{sec:theory}

In this section we briefly describe the formalism proposed by \cite{dl99} and
the method that we use to estimate bias from galaxy counts.
The key step is the procedure to estimate the galaxy PDF,  $P(\delta_g)$, 
from the measured probability of galaxy counts in cells, $P(N_{\it g})$. We review some 
of the techniques proposed to perform this crucial step and describe in detail the technique used in this work.

\subsection{Stochastic non-linear bias}

 \cite{dl99} proposed a probabilistic approach to galaxy bias in which non-linearity and stochasticity 
 are treated independently.  In this framework, galaxy bias is described by the conditional probability of 
 galaxy over-density,
 $\delta_g$,  given the mass over-density  $\delta$: $P(\delta_g|\delta)$. 
 Both quantities are  smoothed on the same scale and treated as random fields.  
If biasing is a local process then  $P(\delta_g|\delta)$ fully characterises 
galaxy bias.  
Key quantities formed from the conditional probability are the mean biasing function
\begin{equation}
b(\delta) \delta \equiv \langle \delta_g | \delta \rangle = \int P(\delta_g | \delta) \delta_g d\delta_g\; ,
\label{eq:mean}
\end{equation}  
and  its non-trivial second-order moments
 \begin{equation}
 \hat{b} \equiv \frac{\langle b(\delta) \delta^2 \rangle}{\sigma^2}  \ \ \ \ \ \ \ \ \ 
 \tilde{b}^2 \equiv \frac{\langle (b(\delta) \delta)^2 \rangle }{\sigma^2} \; ,
 \label{eq:2ndorder}
 \end{equation}
where $\sigma^2 \equiv \langle \delta^2 \rangle$ is the variance of the mass over-density field
on the scale of  smoothing.
The quantity $\hat{b}$ represents the slope of the linear regression
of $\delta_g$ against $\delta$ { and is the natural generalisation of the linear bias parameter}.
The ratio $\tilde{b}/\hat{b}$ { quantifies the deviation of the mean biasing function from a straight line}.
It measures the non-linearity of the mean biasing relation and, in
realistic cases, is close to unity.
In the limit of linear and deterministic bias, the two moments $\hat{b}$ and $\tilde{b}$  coincide with the (constant)
mean biasing function $b(\delta)=b_{\rm LIN}$, where $b_{\rm LIN}$ is the familiar linear bias parameter.
We note that $\hat{b} $ is sensitive to the mass variance and scales as $\hat{b} \propto \sigma^{-1}$.
On the contrary, the moments' ratio is very insensitive to it, 
$\tilde{b} /\hat{b}\propto \sigma^{0.15} $  \citep{Sigad00}. These scaling relations are used 
in Section~\ref{sec:results}
to  compare results obtained assuming different values of $\sigma_8$.
{  There are other useful parameters related to galaxy bias that can be  measured from the data. 
One is the ratio of variances $b_{\rm var}\equiv  (\sigma_g/\sigma)^2 $
in which $\sigma_g$ is measured from counts in cells and $\sigma$ depends on the 
assumed cosmological model.
Another quantity is the inverse regression of $\delta$ over $\delta_g$, $b_{\rm inv}\equiv
\sigma_g^2/\langle \delta_g \delta \rangle $ that requires an estimate of the galaxy  and the mass density fields
 \citep{Sigad98}.
In the case of non-linear deterministic bias these quantities differ from $ \hat{b}$.
Specifically, if the non-linearity parameter  $\tilde{b}/\hat{b}$ is 
larger (smaller) than unity then they are biased high (low) with respect to 
 $\hat{b}$  \citep{dl99}}.
 
 In this paper we focus on the $\hat{b}$ parameter, a choice that
 allows us to compare our results with those of K11 (but not with M05, in which the focus is instead on 
 $\tilde{b} $). Fortunately, as we shall see, the small degree of non-linearity makes these two choices 
 almost equivalent.

If bias is deterministic, then it is fully characterised by the mean biasing function $b(\delta) \delta$.
However,  we do not expect this to be the case since galaxy formation and evolution are regulated by complex 
physical processes that are not solely determined by the local mass density. Therefore, for a given
value of $\delta$ there is  a whole distribution of $\delta_g$ about the mean $b(\delta) \delta$. 
{This scatter,
often referred to as bias stochasticity, is contributed by two sources:  shot noise due to the discrete sampling of a continuous underlying density field  and 
those astrophysical processes relevant to the formation and evolution of galaxies 
that do not depend (solely) on the local mass density.

Previous studies \citep{branchini01, marinoni05, viel05, kovac011} that, like this one, used the galaxy 
1-point PDF to recover the biasing function
ignored the impact of stochasticity and assumed a deterministic bias.
We aim to improve the accuracy of the bias estimator by taking 
 bias stochasticity into account and we do this by assuming that
 shot noise is the only source of stochasticity.
This simplifying assumption can be justified theoretically by both numerical and analytic arguments.
Numerical experiments in which semi-analytic galaxies 
are used to probe the mass density field in samples mimicking
SDSS  \citep{sp04} (see Figs. 11 and 16) and 2MRS \citep{nusser014} (see Fig. 1), i.e.
two surveys with galaxy number densities similar to that of VIPERS,  
do indeed show that shot noise is the dominant source of scatter.
More specifically, Poisson noise accounts for the scatter in the 
 $\delta_g$ versus $\delta$ relation except at large over-density 
where the  relation is over-dispersed.
Analytic arguments  in the framework of the halo model also confirm that the main 
source of stochasticity is shot noise with the halo-halo scatter providing a significant contribution 
for faint objects alone \citep{Cacciato}.
Assessing the impact of this shot noise only assumption is not simple, but some arguments can be made to 
quantify the systematic effect of underestimating stochasticity.

An upper limit can be obtained when stochasticity is ignored altogether. In the case of linear and stochastic bias, for example, $\tilde{b}$ and $\hat{b}$ would be equal
whereas $b_{\rm inv}$ would be systematically larger by about 10\%  \citep{Somerville}. 
The more realistic case of a non-linear and stochastic bias was considered by
\cite{Sigad00} using numerical simulations again. In this case, the effect of 
ignoring stochasticity  is that of overestimating both  $\tilde{b}$ and $\hat{b}$.
The amplitude of the effect depends on both the cosmological model assumed and the scale
considered. To obtain estimates relevant to our analysis we repeated the \cite{Sigad00} test
in  Section~\ref{sec:galpdf}. The results, which we anticipate here, 
indicate that $\tilde{b}$ and $\hat{b}$ are overestimated by 8(4)\% on a scale of 4(8) \hmpc.
As for the ratio, $\tilde{b} /\hat{b}$  we also confirm that it is remarkably insensitive to stochasticity and, as 
expected, to the model adopted \citep{Sigad00}. 

Analyses of the datasets may also constrain  the size of the  effect. 
Galaxy clustering, higher order statistics, or gravitational lensing generally indicate that galaxy bias 
cannot be linear and deterministic. However, as we anticipated in the introduction, 
it is not possible to disentangle the effects induced by non-linearity and stochasticity, except for the
case of relative bias between two types of tracers.
With respect to this, the largest stochasticity $\sigma_b/\hat{b}=0.44$ so far was measured by
 \cite{Wild}. If ignored, this would induce a systematic 
error of  $\sim 20 \%$ on the relative $\hat{b}$ moments. 

 Overall, the  variety of evidence
indicates that if stochasticity is ignored then $\sigma_b$ and $\hat{b}$
are overestimated by  10-20 \%, whereas their ratio is unaffected.  
However,  we stress that  in our work stochasticity is, at least in large part, taken into account.
Therefore, we expect that our assumption that shot noise is the only source of bias stochasticity
generates systematic errors well below the 10 \% level.
}

\subsection{Direct estimate of  $b(\delta) \delta$}
\label{sec:direct}

Under the hypothesis that bias
is  deterministic and monotonic  the mean biasing function, $b(\delta) \delta$,
can be estimated  by comparing the PDFs of the mass and of the galaxy over-density.
We let  $C(\delta)\equiv P(>\delta)$ and $ C_g(\delta_g)\equiv P(>\delta_g)$ be the  cumulative probability distribution functions [CDFs]
obtained by integrating the two PDFs.
Monotonicity guarantees that the ranking of the fluctuations
$\delta$  and $\delta_g$ is preserved and $b(\delta) \delta$
can be obtained by equating the two CDFs at the same percentile,
\begin{equation}
 b(\delta)\delta=C_g^{-1}(C(\delta)) \; ,
\label{eq:inversions}
\end{equation}
where $C_g^{-1}$  indicates the inverse function of $C_g$.

Equation~\ref{eq:inversions} provides a practical recipe
to estimate galaxy  bias from observed counts in cells of a given size.
It requires three ingredients: the galaxy over-density $\delta_g$, its PDF, and that of $\delta$.
$\delta_g$ can be estimated from galaxy counts in cell, $N_g$ as
\begin{equation}
1+\delta_g=N_g/\langle N_g \rangle \, ,
\label{eq:deltan}
\end{equation}
where $\langle N_g \rangle$ represents mean over all counts.
From Eq. \ref{eq:deltan} one can form the galaxy PDF, $P(\delta_g)$ and 
the count probability $P(N_g)$. 
The biasing function can then be obtained by 
comparing $C_g(\delta_g)$ with a model $C(\delta)$.

This simple bias estimator  has been used by several authors
 \citep{Sigad00, branchini01, marinoni05, viel05, kovac011}.
It is potentially affected by several error sources that should be systematically investigated.
The first error source is shot noise that affects the estimate of $\delta_g$ from $N_g$.
Shot noise induces stochasticity in the bias relation in contrast with the 
hypothesis of deterministic bias. 
Stochasticity affects the estimate of   $b(\delta) \delta$ from Equation~\ref{eq:inversions}, especially
at large values of $\delta_g$, where the CDF flattens and the evaluation of the 
inverse function $C_g^{-1}$ becomes noisy.
A second issue is the mass PDF for which no simple theoretical model is available.
The last error source is redshift distortions.
Galaxy over-densities are computed using the redshift of the 
objects rather than distances. This induces systematic differences 
between densities evaluated in real and redshift space \citep{kaiser87}.

All these issues potentially affect the estimate of galaxy bias and should be properly quantified and accounted for. 
In the next section, we review some existing estimators 
designed to minimise the impact of the shot noise and propose a new estimator 
that we apply in this paper.
{ We investigate the performance of this new strategy  in Section~\ref{sec:esources}.}

\subsection{ From $P(N_g)$ to $P(\delta_g)$... }
\label{sec:PDFestimator}

The  probability of galaxy counts, $P(N_g)$, can be expressed as
\begin{equation}
P(N_g)=\int_{-1}^{+\infty}P(\delta_g)P(N_g|\delta_g)d\delta_g \, ,
\label{eq:convolution}
\end{equation}
where the conditional probability function $P(N_g|\delta_g)$ 
specifies the way in which discrete galaxies sample the underlying, continuous field.
The common assumption that galaxies 
are a local Poisson process implies that 
\begin{equation}
P(N_g | \delta_g)=\frac{\left[\langle N_g\rangle (1+\delta_g)\right]^{N_g} e^{-\langle N_g \rangle(1+\delta_g)}}{N_g!} \, .
\label{eq:kernel}
\end{equation}
{ The Poisson model provides a good match to numerical experiments except at large densities { where}
a negative binomial distribution seems to provide a better fit  \citep{sheth15,som01,CM02}.
In this work we adopt the Poisson model. { However, }
 different forms for $P(N_g|\delta_g)$ { could} be considered as well.}

The following strategies have been proposed to estimate $P(\delta_g)$ from $P(N_g)$ using Equation \ref{eq:convolution}:

\begin{itemize}

\item {\it Richardson-Lucy deconvolution}. \cite{sp04} proposed this iterative, non-parametric method to
reconstruct $P(\delta_g)$ by comparing the observed $P(N_g)$ to that computed from 
Eq.~\ref{eq:convolution} at each step of the iteration, starting from an initial guess for $P(\delta_g)$.

 \item {\it Skewed lognormal model fit}. This parametric method was also implemented by \cite{sp04}.
 In this approach one assumes a skewed lognormal form for $P(\delta_g)$ and then determines the four free parameters
 of the model by minimising the difference between Eq.~\ref{eq:convolution} and the observed $P(N_g)$. 
 
\item {\it Gamma expansion} [$\Gamma_E$]. 
Among the various forms proposed to model the galaxy PDF, the  Gamma expansion, 
 defined by expanding the Gamma distribution on a basis of  Laguerre polynomials   \citep{md10}   captures the essential features of the galaxy density field.
 The expansion coefficients  directly depend on the  
 moments of the observed counts.  Because of this, the full shape of the  
 galaxy PDF can be recovered directly from the observed $P(N_g)$ with no need to 
 integrate Eq.~\ref{eq:convolution}.
 \end{itemize}

 \cite{sp04} have tested the ability of the first two methods  in reconstructing the PDF of 
 halos and mock galaxies obtained from $N$-body simulations. They showed
 that a successful reconstruction can be obtained  when the sampling is 
  $\langle N_g \rangle \ge 0.1$; safely a factor 3 smaller than the smallest 
mean galaxy density in our VIPERS subsamples.
\cite{Bel2} extensively tested the $\Gamma_E$-method
 and  showed, using the same mock catalogues as in this paper,
 that this method reconstructs the PDF of a VIPERS-like galaxy distribution 
 with an accuracy that is superior to that of the other methods.
 This comes at the price of discarding counts in cells that overlap the observed 
 areas by less than 60 \%, which is  a constraint that further {reduces}
  deviations from the Poisson sampling hypothesis.

\begin{figure}
\includegraphics[width=0.55\textwidth]{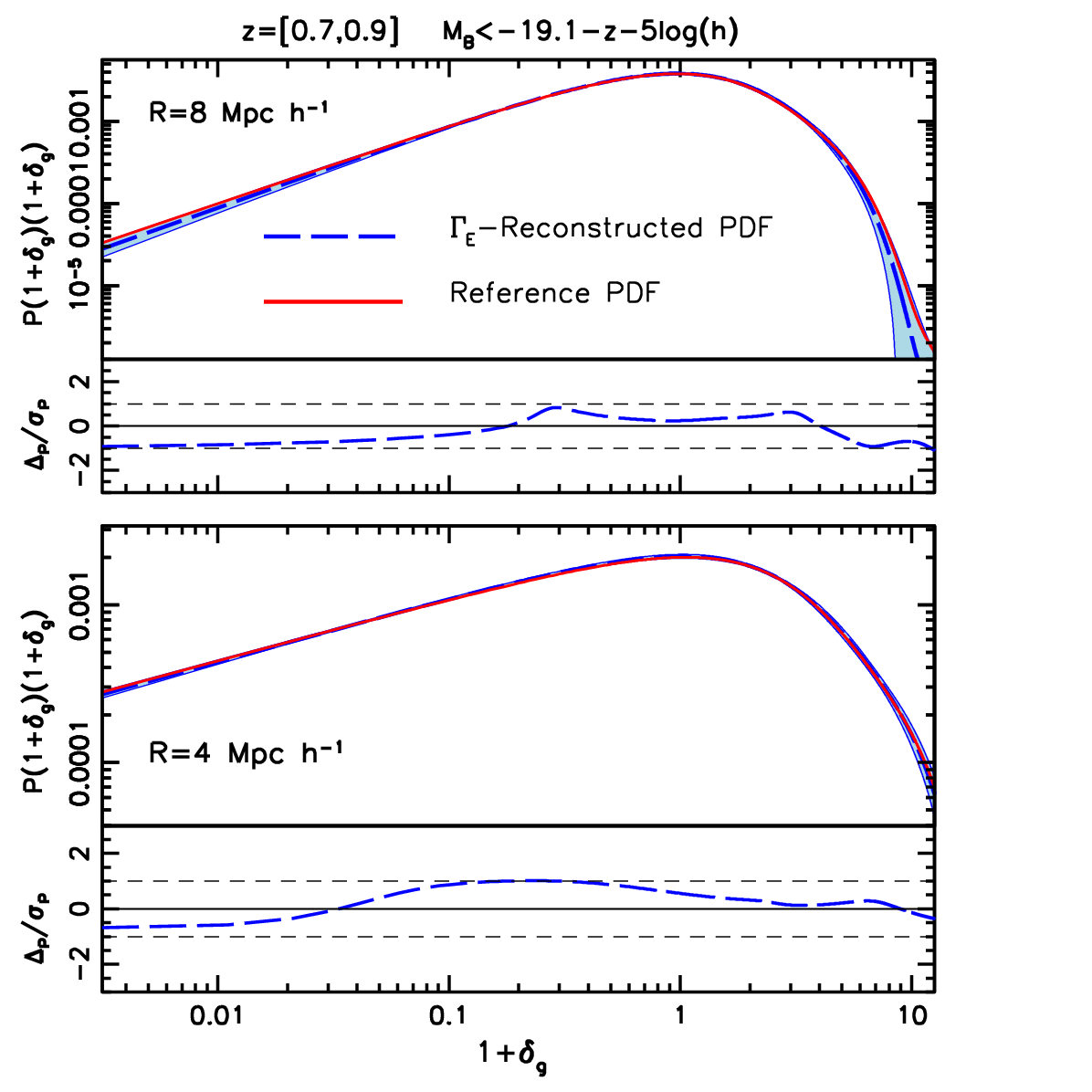}
\caption{
Reconstructed PDF of the mock VIPERS galaxies measured in cells of 
$R=4$ \Mpch { (top)} and $R=8$ \Mpch (bottom).
The blue solid curve represents the reference galaxy PDF obtained by averaging over the PDFs reconstructed from the 
{\it Parent} mocks using the $\Gamma_E$ method.  The blue dashed curve shows the  average PDF reconstructed from 
the {\it Realistic} mocks using the same method.
The blue shaded region represents the 1-$\sigma$
scatter among the 26 {\it Realistic} mocks. 
We plot $P(1+\delta_g)(1+\delta_g)$ to highlight the performance of the 
reconstruction at high and low over-densities. We note the different Y-ranges in the two panels. 
{ The bottom panels in each plot show the difference $\Delta_p$ between the reconstructed and reference PDFs in units of  the random error $\sigma_p$. Horizontal, dashed lines indicate systematic errors equal to 1-$\sigma_p$ random uncertainties.}
}
\label{fig:pdftest}
\end{figure}

To illustrate the performance of the $\Gamma_E$-method we plot, in Figure~\ref{fig:pdftest},
the galaxy PDFs $\Gamma_E$-reconstructed from the { 26} {\it Realistic} mock VIPERS subsamples
with galaxies brighter than $M_{\rm B}=-19.1-z-5log(h)$ in the range $z=[0.7,0.9]$. 
The blue dashed curve represents the mean among the mocks and the blue band the 1-$\sigma$ scatter.
The scatter for cells of $R=8$ \Mpch is larger than for $R=4$ \Mpch and is driven by the limited 
number of independent cells rather than sparse sampling.


The reconstruction is compared with the ``reference'' PDF (solid, red line) obtained by averaging over the PDFs 
reconstructed,  with the same $\Gamma_E$ method,  from  the {\it Parent} mock catalogues.
We regard this as the ``reference'' PDF since, as shown by  \cite{sp04} and checked by us, 
when the sampling is {dense}, all the above reconstruction methods
recover the PDF of the mass, $P(N_g)$ and the mean biasing function very accurately.
In the plot we show $P(1+\delta_g)(1+\delta_g)$ to  highlight the low- and high-density tails, where
the reconstruction is more challenging.
The reconstructed PDF underestimates the reference PDF in the low- and high-density tails 
and overestimates it at $\delta \sim 0$. 
Systematic deviations in the low- and high-density tails are to be expected since the
probability of finding halos, and therefore mock galaxies, in these regimes significantly deviates from the probability 
expected for a Poisson distribution.
However, these differences are well within the 1-$\sigma$ uncertainty strip { as shown in the bottom panels of each plot}.

The $\Gamma_E$ method used to reconstruct the galaxy PDF from discrete counts is implemented as follows:

\begin{itemize}

\item We consider as the input dataset  one of the volume-limited, luminosity complete subsamples
listed in Table~\ref{tab:cats}.  The position of each object in the catalogue is specified in redshift space, i.e.
by its angular position and measured spectroscopic redshift.


\item Spherical cells are thrown at random positions within the surveyed region.
We consider cells with radii $ R=4, \, 6, $ and  8  \hmpc. The smallest radius
is set to guarantee $\langle N_g \rangle \geq 0.3$.
The largest radius is set to have enough cell statistics to sample 
$P(N_g)$ at large $N_g$.
We only consider cells that overlap by more than 60 \% with the observed areas. 
This constraint reduces deviations from Poisson statistics  \citep{bel14}.
{ Counts in the partially overlapping cells are weighted by the fraction $f$ of the
surveyed volume in the cell: $N_g/f$.}
The probability function $P(N_g)$ is then computed from the counts frequency distribution.

\item We use the measured  
$P(N_g)$ and its moments to model the galaxy PDF 
with the $\Gamma_E$ method that we compute using 
all factorial moments up to the sixth order.

\end{itemize}

\subsection{....and from $P(\delta_g)$ to $b(\delta) \delta$. }
\label{sec:procedure}

To estimate the mean biasing function from  the galaxy PDF, we solve  Equation ~\ref{eq:inversions}.
{ To do so, we assume that shot noise is the main source of stochasticity and} 
 that a reliable model for the mass PDF is available.
Despite its conceptual simplicity, this procedure requires
several non-trivial steps that we describe below. The uncertainties introduced in each step
 are estimated in the next section. The procedure is as follows:

\begin{itemize}

\item { We start from the galaxy PDF estimated from the measured $P(N_g)$, as { described} in the previous {section.}

\item We assume a model PDF for the mass density field in redshift space. Rather than adopting some approximated, analytic model, we 
measure the mass PDF directly from a dark matter only $N$-body simulation 
with the same characteristics and cosmological model as the Millennium run \citep{springel05},
that
 is not based on the same model used to build the HOD-mock VIPERS catalogues.
The use of an incorrect mass PDF is yet another possible source of systematic errors that we quantify in 
Section~\ref{fig:masspdf}. However, this error is expected to be small
since $\hat{b}$  and $\tilde{b}$ are { mainly}  sensitive to $\sigma$ 
and their ratio  is largely independent of the underlying cosmology
\citep{Sigad00}.

\item { After computing the cumulative distribution function from the mass and galaxy PDFs,} we 
use Eq.~\ref{eq:inversions} to estimate the mean biasing function.

\item { We determine the maximum over-density $\delta_{\rm MAX}$ at which the reconstructed mean biasing function
can be considered reliable. 
To estimate $\delta_{\rm MAX}$ we compare the measured $P(N_g)$ 
with the estimated $P(N_g)$ following} the procedure described in Section~\ref{sec:dmax} .}


\item { We estimate the second-order moments $ \hat{b}$ and  $\tilde{b} $ and their ratio by integrating over all $\delta$ up to $\delta_{\rm MAX}$
\begin{eqnarray}
\nonumber
\hat{b}&=& \sigma^{-2} \int_{-1}^{\delta_{\rm MAX}} b(\delta)\delta^2 P(\delta) \  d\delta \, ,
\\
\tilde{b}^2&=&\sigma^{-2} \int_{-1}^{\delta_{\rm MAX}} (b(\delta)\delta)^2 P(\delta) \ d\delta \;.
\label{eq:bhatdelta}
\end{eqnarray}
and test the robustness 
of the result with respect to the choice of  $\delta_{\rm MAX}$.
 }

\end{itemize}


\section{Error sources}
\label{sec:esources}

In this Section we review all possible sources of uncertainty that might affect 
the recovery of the biasing function and assess their amplitude using  mock catalogues.
In this process we need to consider a reference biasing function to compare with the results of the reconstruction.
This could be estimated directly from the distribution of the dark matter particles and mock galaxies within the simulation box.
{ However,  we use the mean biasing function obtained from the  {\it Parent} mocks
as reference. We justify this choice as follows.
First, \cite{sp04} showed that when the sampling is dense 
both the Richardson-Lucy and  the skewed lognormal fit methods recover the mean biasing function 
with high accuracy. Second, in Section~\ref{sec:PDFestimator} we found that when the sampling is dense
the $\Gamma_E$ method accurately recovers the mean biasing function in the {\it Parent} mocks.}

\subsection{Sensitivity to the galaxy PDF reconstruction method}
\label{sec:galpdf}

\begin{figure}
\includegraphics[width=0.55\textwidth]{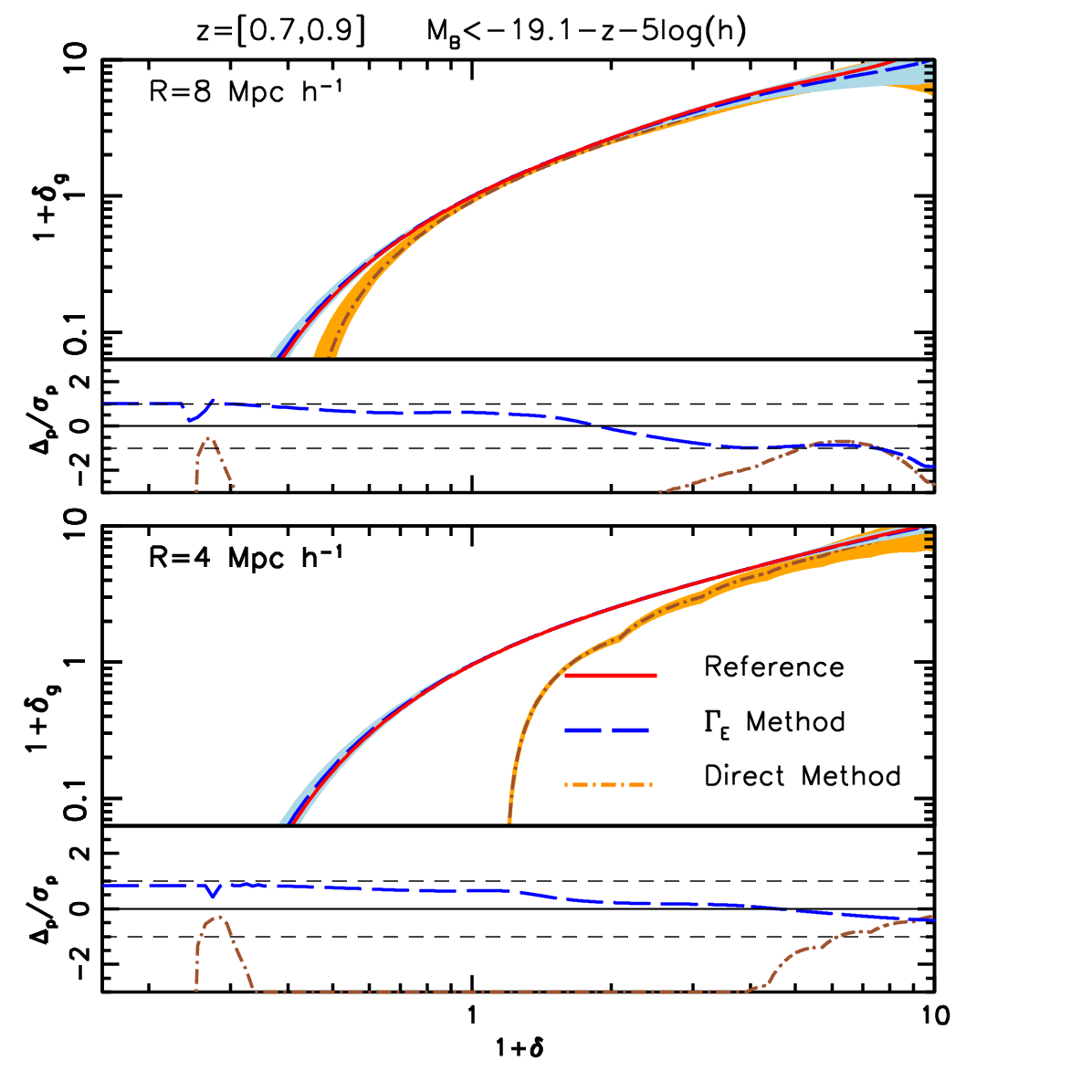}
\caption{
Mean biasing function of mock VIPERS galaxies computed from counts 
in cells of  $R=4$ \Mpch (bottom panel) and $R=8$ \Mpch (top panel).
The magnitude cut and redshift range of the mock VIPERS subsample, indicated in the plot,
are the same as Figure~\ref{fig:pdftest}.
Solid red curve: reference biasing function { obtained from the {\it Parent} mock catalogues}. Blue dashed curve and 
blue-shaded region: average value and 2-$\sigma$ scatter of the biasing function 
reconstructed from the {\it Realistic} mocks using the $\Gamma_E$ method.Brown dot-dashed curve and orange-shaded band: average value and 2-$\sigma$ scatter of the biasing function 
reconstructed from the {\it Realistic} mocks using a 'direct' estimate of the galaxy PDF.
{ Bottom sub-panels: difference $\Delta_p$ between the reconstructed and reference PDFs in units of the random error $\sigma_p$. Dashed lines indicate systematic errors equal to 1-$\sigma_p$ random errors.}
}
\label{fig:directtest}
\end{figure}

Most of the previous estimates of the mean biasing function 
did not attempt to account for shot noise directly. This choice can 
hamper the recovery of $b(\delta) \delta$ when the sampling is sparse.
To estimate errors induced by ignoring shot noise and quantify the benefit of using 
the $\Gamma_E$ method 
we  compared the biasing functions reconstructed using both procedures. 
The result of this test is shown in Figure~\ref{fig:directtest}.
The red curve represents the reference biasing function obtained by averaging 
over the {\it Parent} mocks. In each mock
the biasing function was estimated
from the galaxy PDF using the $\Gamma_E$ method.
The blue dashed curve represents the same quantity estimated from the { 26} {\it Realistic} mocks
using the $\Gamma_E$ method.
The blue band represents the 2-$\sigma$ scatter. 
For negative values of $\delta_g$ the reconstructed biasing function is below the reference biasing function,
but the trend is reversed for  $\delta_g>0$, reflecting the mismatch between the reconstructed and 
reference PDFs in Figure~\ref{fig:pdftest}. 
{ The discrepancy however, is mostly within the 2-$\sigma$ scatter (horizontal dashed line in the bottom
sub-panels).}
On the contrary, the biasing function obtained from the ``direct'' 
estimate of $\delta_g$ (brown dot-dashed curve and {the} corresponding  2-$\sigma$ scatter, orange band)
is significantly different from the reference function.  The discrepancy increases 
at low densities and for small spheres, i.e. when the counts per cell decrease and the shot noise 
is large.

\subsection{Sensitivity to the mass PDF}
\label{sec:masspdf}

Another key ingredient of the mass reconstruction is the mass PDF. 
In principle this quantity  could be obtained from galaxy peculiar velocities or gravitational lensing.
However, in practice,  errors are large and would need to be averaged out over scales much larger than 
the size of the cells considered here. For this reason we need to rely on theoretical modelling.
\cite{coles91} and \cite{kofman94} found that the mass PDF can be
approximated by a lognormal distribution and this model
 was indeed adopted 
in previous reconstructions of the biasing function (e.g. M05, \cite{Wild}, K11).

\begin{figure}
\includegraphics[width=0.55\textwidth]{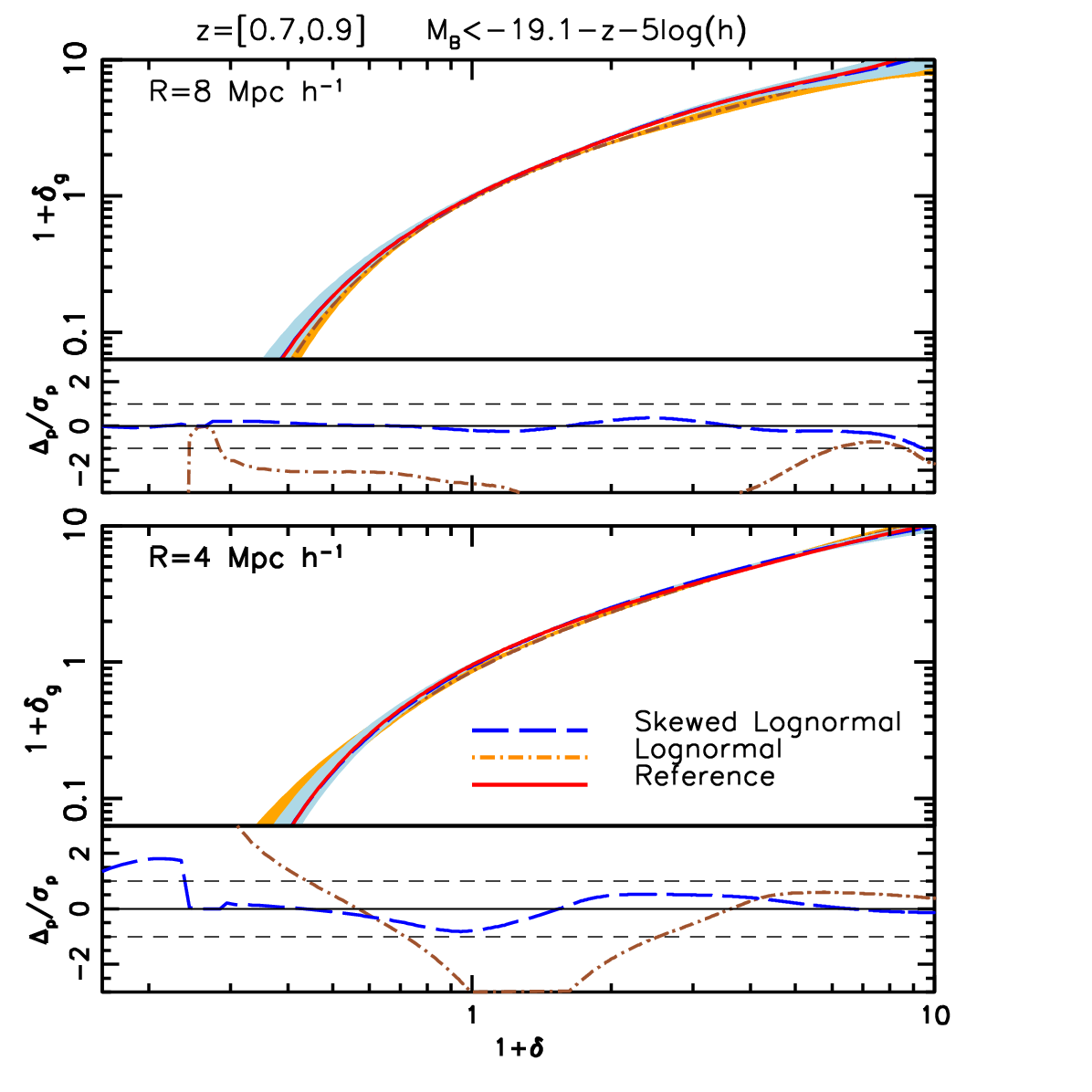}
\caption{Solid red curve: reference mean biasing function of Fig.~\ref{fig:directtest}
computed using the mass PDF from $N$-body simulations.
Brown dot-dashed curve and orange band: biasing function obtained using a lognormal fit to the mass 
PDF and 2-$\sigma$ scatter from the mocks.
Blue dashed curve and blue band: biasing function obtained using a skewed lognormal fit to the mass 
PDF and  2-$\sigma$ scatter from the mocks.
{ Bottom panels: difference $\Delta_p$ between the reconstructed and reference PDFs in units of the random error
$\sigma_p$. Dashed lines indicate systematic errors equals to 1-$\sigma_p$ random errors.}
}
\label{fig:masspdf}
\end{figure}

However, the lognormal approximation is known to perform poorly in the high- and low-density tails and for 
certain spectra of density fluctuations.
An improvement over the lognormal model is represented by 
the skewed lognormal distribution \citep{colombi94}. 
This model proved to be an excellent approximation
to the PDF of the dark matter measured from $N$-body experiments
over a wide range of scales and of over-densities \citep{ueda96}.
The impact of adopting either  model for the mass PDF 
can be appreciated in Fig.~\ref{fig:masspdf}.
The solid red curves represent the same biasing functions
shown in Fig.~\ref{fig:directtest} obtained from the galaxy PDFs of the {\it Parent } mocks and
from a mass PDF obtained directly from an $N$-body simulation 
with the same cosmological parameter and size as the 
Millennium simulation using the output corresponding to $z=0.8$.
As in the previous test, we consider the red solid curve as the reference biasing function.
The brown dot-dashed curve shows the mean biasing function reconstructed assuming a 
lognormal model for the mass PDF, i.e. a lognormal fit to the PDF measured from the 
$N$-body simulation. The curve represents the average among { 26} mocks and the 
orange band  is the 2-$\sigma$ scatter.  
{ For $R=8$ \hmpc, the biasing function is systematically below the reference
whereas for $R=4$ \hmpc is above 
the reference at both high and low densities. The mismatch is very large and significantly
exceeds the 1-$\sigma$ scatter (bottom sub-panels). The skewed lognormal model 
(blue dashed curve) performs significantly better with differences well below 1-$\sigma$ 
except at very negative $\delta$ values.}

We conclude that, for the practical purpose of reconstructing galaxy bias, the 
mass PDF measured from $N$-body data and a skewed  lognormal fit perform equally well.
The main advantage of using the latter would be the possibility of determining the four parameters of the fit experimentally. Since, however, the parameters are poorly constrained by observations,
we decided to adopt the mass PDFs  from $N$-body  simulations.
{ This choice introduces a dependence on the cosmological model,
however, that 
is mostly captured by one single parameter,  $\sigma$,
for which  $\hat{b}$ and $\tilde{b}$ { exhibit a linear dependent}.}
With respect to this, the mass PDF used to obtain the biasing functions in 
Figure~\ref{fig:masspdf} is not the {\it true} mass PDF since it is obtained 
from an $N$-body simulation that uses a cosmological model 
that is different from the model used to produce the mock catalogues.
{ We did this on purpose { to mimic} the case of the real analysis 
for which the underlying cosmological model is not known.}

\subsection{Sensitivity to redshift distortions}
\label{sec:zditosrtions}

Galaxy positions are measured in redshift space, i.e. using the observed redshift to estimate the
distance of the objects. The presence of peculiar velocities 
induces apparent radial anisotropies in the spatial distribution of galaxies and, as a consequence, modifies
the local density estimate and their PDF  \citep{kaiser87}. However, our goal is to 
reconstruct the mean biasing function in {real space} without redshift distortions.
{ Considering the difficulties and uncertainties in determining the galaxy PDF in real space,
one could { instead consider} the galaxy and mass PDFs both measured in redshift space
under the assumption}
that peculiar velocities induce similar distortions in 
the spatial distribution of both dark matter and galaxies so that they cancel out when 
estimating the mean biasing relation from Eq.~\ref{eq:inversions}.
In the limit of the Gaussian field, linear perturbation theory and no velocity bias, the cancelation is exact.
However, non-linear effects have a different impact on the mass and galaxy density fields and 
induce different distortions in their respective PDFs.
To assess  the impact of these  effects  we compared the mean biasing function 
of mock galaxies reconstructed
from PDFs estimated in real and redshift space.

\begin{figure}
\includegraphics[width=0.56\textwidth]{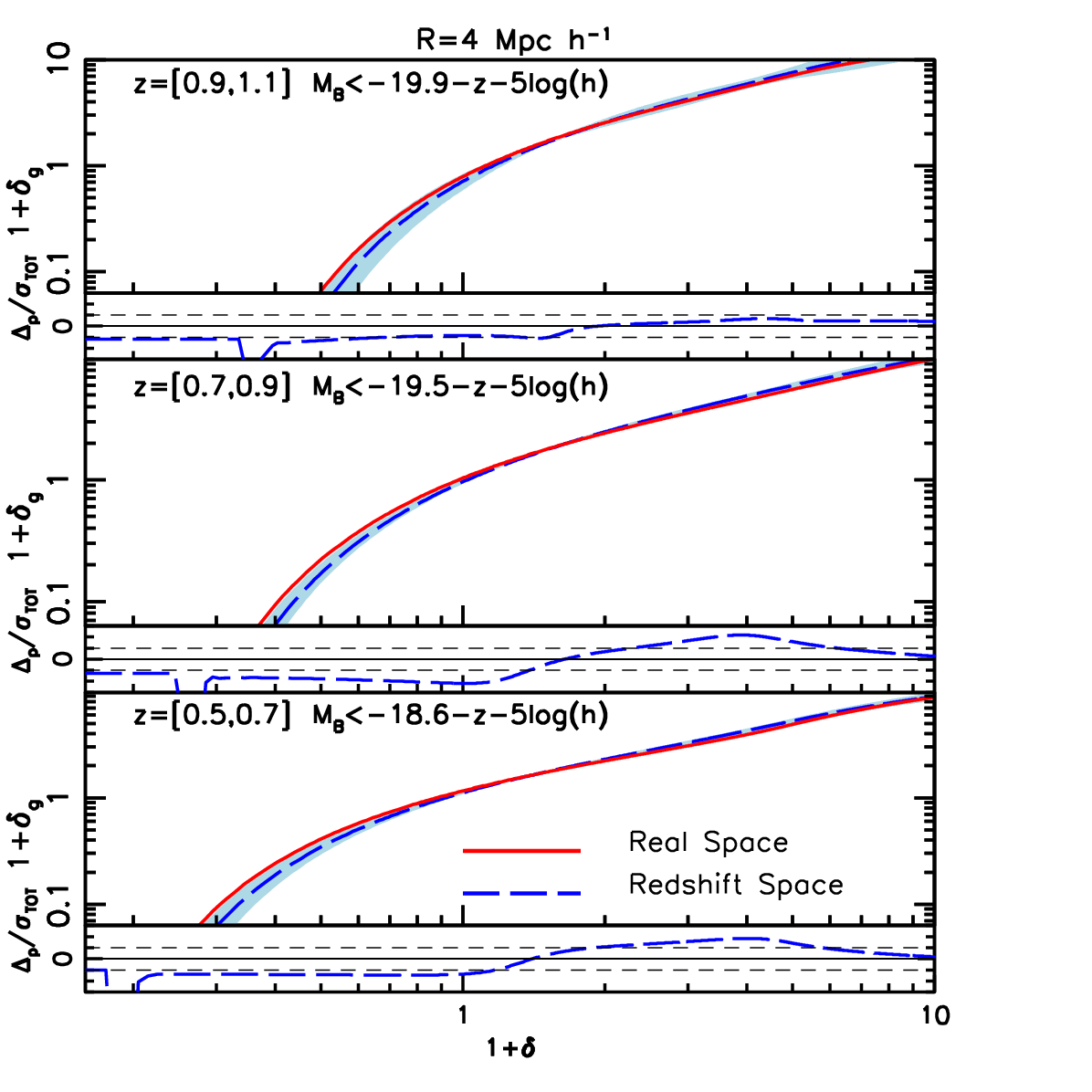}
\caption{ 
Mean biasing function estimated in real space (solid, red curve) and redshift space (dashed blue curve and 
its 2-$\sigma$ uncertainty band). 
Counts are performed in spherical cells with a radius of 4 \Mpch.
The luminosity cut and the redshift range is indicated in each panel. 
 The width of each band represents the scatter among mocks.
 { In the bottom part of each plot we show the difference $\Delta_p$ between the reconstructed and reference PDFs in units of 
$\sigma_{\rm TOT}$, where $\sigma_{\rm TOT}$ accounts for the {\it rms} scatter in both the real- and redshift-space mocks.
 Dashed lines indicate where systematic errors equal to 1-$\sigma_{\rm TOT}$ random errors.}
}
\label{fig:ztest}
\end{figure}

The results are shown in  Figure~\ref{fig:ztest}. The solid red curve represents the mean biasing function
of galaxies in the {\it Realistic} mock catalogues estimated using the PDFs of galaxies and mass in real space.
The blue dashed line shows the same function estimated in redshift space. Both curves are obtained by averaging
over the  26 mocks and the blue band represents the 2-$\sigma$ scatter in redshift space.
The redshift space biasing function underestimates the true biasing function
in low-density regions  
and overestimates it at high densities, i.e. in the presence of highly non-linear flows.
{ The difference is systematic but its amplitude is within the 2-$\sigma$ random errors
estimated by { adding in quadrature the scatter among mocks in real and redshift space}
(bottom panels in each plot)}.
{The biasing functions 
 shown in Figure~\ref{fig:ztest} represents a demanding test in which we consider the smallest cells of 4 \hmpc
where deviations from linear motions are larger. The discrepancy
decreases if the size of the cell increases.}

\begin{figure}
\includegraphics[width=0.56\textwidth]{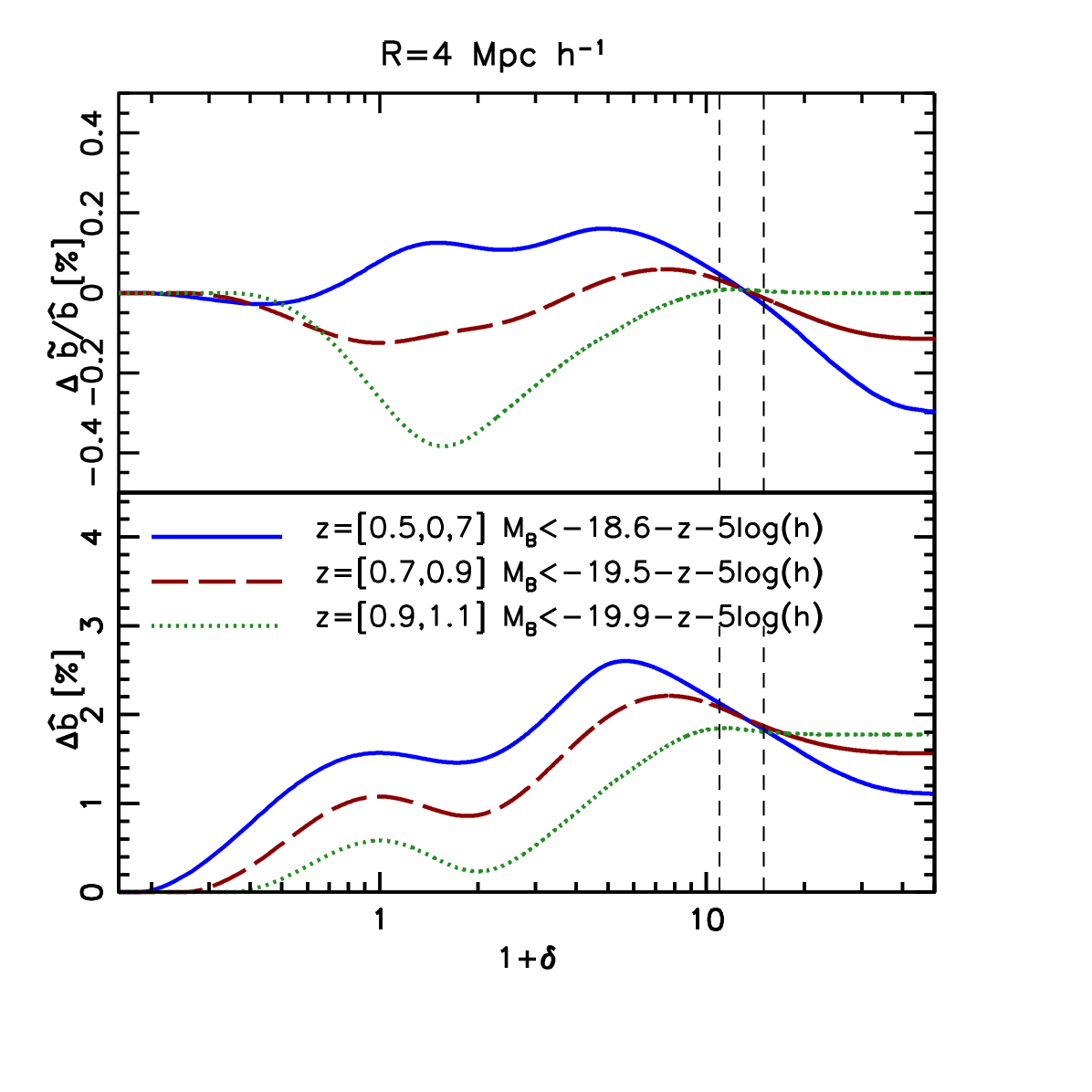}
\caption{ 
{  Bottom panel: per cent difference between the  $\hat{b}$ values estimated in redshift space and in real space
using spherical cells with a radius of 4 \Mpch as a function of $1+\delta_{\rm MAX}$ (see Eq.~\ref{eq:bhatdelta}).
The different curves refer to different redshift shells and magnitude cuts, as indicated in the plot.
Upper panel: per cent difference in the estimated non-linear parameter $\tilde{b}/\hat{b}$. 
Vertical dashed lines are drawn in correspondence to
the $\delta_{\rm MAX}$  values at which systematic errors are computed and listed in Table~\ref{tab:errors}.   
 }
}
\label{fig:ztest2}
\end{figure}

{ These systematic differences induce errors in the estimated moments  $\hat{b}$ and $\tilde{b}$. 
To quantify the effect we computed the moments as a function of $\delta$ (i.e. by varying $\delta_{\rm MAX}$
in Eq.~\ref{eq:bhatdelta}) both in real and redshift space. 
The results are shown in Figure~\ref{fig:ztest2}. The plots show the
per cent difference between the 
moments measured in redshift versus real space. 
The panels and curves refer to the same redshift bins and magnitude cuts as in Figure~\ref{fig:ztest}. 
Systematic errors induced by redshift distortions are $\sim 2$ \% for  $\hat{b}$ and for  $\tilde{b}$ (not shown)
and one order of magnitude smaller for $\tilde{b}/\hat{b}$. 
They provide the main contribution to the total systematic errors listed in Table~\ref{tab:errors}
and are of the same size, although somewhat smaller than the random errors.  

Considering the absolute and relative size of these errors, 
we perform our analysis in redshift space.} 

\subsection{Error estimate}
\label{sec:errors}

Different sources of errors affect the recovery of the biasing function. One error source is cosmic variance due to the finite volume of 
the sample. This source dominates the error budget of the M05 and K11 analyses.

The other sources are the shot noise induced by discrete sampling and the limited number of independent 
cells used to build the probability of galaxy counts $P(N_g)$. 
In the VIPERS survey, which is based on a single-pass strategy, 
sparse sampling is more of an issue than in the M05 and K11 cases.
The cumulative effect of the single pass strategy and  colour preselection
reduces the sampling rate to $\sim 35$ \% on average with significant variations across quadrants.
The survey geometry, characterised by gaps and missing quadrants that
occupy $\sim 25$ \% of the would-be continuous field, further dilutes the sampling (we consider cells that overlap
up to $40$ \% with unobserved regions) and  limits the number of independent 
cells that can be accommodated within the survey.
Our PDF reconstruction strategy is designed to minimise these effects that, nevertheless, 
induce random and systematic errors that need to be estimated.
We do this with the help of both the {\it Parent} and {\it Realistic} mock catalogues. The former provide the 
reference mean biasing function. Errors are estimated by comparing the bias function reconstructed  from
the {\it Realistic} mocks to the reference mocks.
The procedure is detailed below and the  estimated errors are {listed} in Table~\ref{tab:errors}.

\subsubsection{Total random error}
\label{sec:toterrors}

To estimate the total random error $\sigma_{\rm RND}$, we proceed as follows.
We reconstruct the mean biasing function in each of the {\it Realistic} mock catalogues, 
compute the average over the {26} mocks and, finally,  estimate the scatter around the mean.
The {\it rms} scatter provides an estimate of the total random error. { All} sources of uncertainties contribute to this error 
{(e.g. cosmic variance, shot noise, and limited number of cells),}
which may affect the recovery of the biasing function. Total random errors for $\hat{b}$ and
$\tilde{b}/\hat{b}$ are listed in columns 6 and 10 of Table~\ref{tab:errors}, respectively.   

\subsubsection{Cosmic variance}
\label{sec:cverrors}
To assess the contribution of the cosmic variance, $\sigma_{\rm CV}$,  to the error budget, we proceed 
as for the estimate of total random errors using, however, the {\it Parent} catalogues rather than 
the {\it Realistic} catalogues. Since errors in the bias reconstruction are mainly driven by discrete sampling and
in the {\it Parent } catalogues the sampling is dense, the {\it rms} scatter among these mocks 
is dominated by cosmic variance. Cosmic variance contributions to errors in $\hat{b}$ and
$\tilde{b}/\hat{b}$ are { shown} in columns 7 and 11 of Table~\ref{tab:errors}, respectively.   
It turns out that the contribution of the cosmic variance is of the same order as that of the sparse sampling
{and}, unlike in the case of M05 and K11, it does not dominate the error budgets.

\subsubsection{Systematic errors}
\label{sec:syserrors}
Following K11, we compute systematic errors, $\sigma_{\rm  SYS}$,
as the average offset of the bias estimates in the 
 {\it Realistic}  and the  {\it Parent} catalogues, i.e. $\sigma_{\rm  SYS}=\langle X_{\it Realistic} -  X_{\it Parent} \rangle$,
 where $X$ is either $\hat{b}$ or $\tilde{b}/\hat{b}$ and
the mean is over the { 26} pairs of mocks.
{ These systematic errors are plotted in the bottom panels of  Fig.~\ref{fig:directtest} (blue, dashed curves).
Their amplitudes at $\delta_{\rm MAX}$ are listed { in}
columns 8 and 12 of Table~\ref{tab:errors}. { These systematic errors} are of the same order as the random errors
and  as the errors induced by redshift distortions discussed in Section~\ref{sec:zditosrtions}.
{ These systematic errors include those induced by redshift distortions. The fact that 
they are of the same order as those discussed in Section~\ref{sec:zditosrtions} indicates that 
they dominate the budget of systematic errors.}

Our systematic errors  are similar to those estimated by K11 (upper part of their Table 2)
from the $z$COSMOS sample, which is significantly small than VIPERS. As these errors do not seem to depend on 
the volume of the survey, we conclude that they can be regarded as genuinely systematic. 
Systematic errors on $\hat{b}$ are on average positive, meaning that 
the mean slope of the reconstructed biasing function typically overestimates the true biasing function.
 For the non-linear {\it bias,} parameter systematic errors are preferentially negative, indicating that 
the reconstruction procedure has the tendency to underestimate
the non-linearity of the biasing function.

\subsubsection{The value of $\delta_{\rm MAX}$}
\label{sec:dmax}

{ 

Our {\it bias} estimator becomes progressively less reliable as the density increases,
for two reasons: first,
the numerical solution to Eq.~\ref{eq:inversions} becomes unstable when the cumulative distribution functions approach unity, i.e.
in correspondence of the high peaks of the mass and galaxy density fields. In this regime, small errors in the estimated 
galaxy PDF propagate into large uncertainties in { $\delta$}; second,
as anticipated in the previous Section, the scatter in the $\delta_g$ versus  $\delta$ relation is larger than Poisson. 
{Our assumption} that Eq.~\ref{eq:kernel}  is valid at all $\delta$ leads to underestimating the high-density tail of the galaxy PDF and, consequently, 
the value of $\hat{b}$. 

Our mock catalogues can be used to {estimate the} first {type} of error, but cannot fully account for the second type of error since 
{our} mock galaxies are sampled from dark matter halos assuming  Poisson statistics. 
{We therefore take the alternative route of reducing} the impact of deviations from Poisson statistics at high densities.
{ We do this by setting a sensible} maximum over density value, $\delta_{\rm MAX}$, at which we compute the bias moments. The value of this threshold is 
computed as follows:
\begin{enumerate}
\item We consider the difference $\Delta P$ between the
`true' $P^t(N_g)$ measured in the  {\it Realistic} mock catalogues and the reconstructed $P^r(N_g)$ estimated through 
Eq.~\ref{eq:convolution}.
\item We search for the first $N_g$ value, $N_{\rm 1}$, at which $\Delta P > 2 \sigma_P$, 
where $\sigma_P$ is the {\it rms} scatter in the mocks.
\item We search for the first $N_g$ value, $N_{\rm 2}$, at which $|\Delta P/P^t(N_g)|>0.5.$
\item We take $N_{\rm MAX}=Min[N_{1},N_{2}]$ and compute the corresponding over-density in galaxy counts
$\delta_{g,{\rm MAX}}=N_{\rm MAX}/\langle N \rangle $.
\item We obtain the corresponding mass over-density $\delta_{\rm MAX}$ from $\delta_{g,{\rm MAX}}$ from the
estimated mean biasing function. 
\end{enumerate}

 The largest over-density at which we search for a solution to Eq.~\ref{eq:inversions} is $\delta_{\rm MAX}$, and this is also the over-density at which
we estimate the bias moments. 
This value is clearly model dependent since it {was estimated from} the VIPERS mocks. An alternative way of setting this threshold 
 would be to look for wiggles in the mean biasing function { measured from real data, i.e. spurious features} induced by instabilities in the 
reconstruction procedure. We found that this second criterion is less stringent as it produces
$\delta_{\rm MAX}$ values larger than using mocks. We decided to adopt a conservative approach and use the $\delta_{\rm MAX}$ 
thresholds estimated with the { first} procedure.

With this criterion we obtain different $\delta_{\rm MAX}$ for the different galaxy subsamples considered in our analysis. { This
limits} our ability to compare results. Since the value of $\delta_{\rm MAX}$ mainly depends on the radius of the cell, we
use one single value for $\delta_{\rm MAX}$ for a given cell size, irrespective of the other parameters used to define the 
subsample. These values, which are listed in 
Table~\ref{tab:errors}, correspond to the minimum $\delta_{\rm MAX}$ among those computed for all subsamples.

All bias parameters presented in our work were computed at these over-density values.
To check the robustness of our results to $\delta_{\rm MAX}$ we also considered a second, less stringent threshold
obtained by taking the {maximum}  value of $\delta_{\rm MAX}$ among those of the various subsamples for a given cell size.
This second set of  $\delta_{\rm MAX}$ {that we denote as}  $\bar{\delta}_{\rm MAX}$, is also listed in Table~\ref{tab:errors}
together with the corresponding estimates for the bias moments (values in parenthesis).

}

\section{Results}
\label{sec:results}

In this section we present the results of our analysis, focusing on the 
dependence of the mean biasing function and its moments on various quantities.
In Sections \ref{sec:mag} and \ref{sec:reds} we explore the 
bias dependence on magnitude and redshifts, respectively. 
In both cases we fix the radius of the cells equal to 6 \hmpc.
The dependence on the cell size  is investigated in Section~\ref{sec:scale}.
Results are summarised in Section~\ref{sec:full}
and listed in Table~\ref{tab:errors}.


\subsection{Magnitude dependence}
\label{sec:mag}

The different solid curves in Figure~\ref{fig:bias_mag} represent the mean biasing function of VIPERS galaxies 
reconstructed from counts in cells of radius 6 \hmpc for different magnitude cuts
 for three  different redshift shells (the three panels). 
We  applied a small { horizontal} offset $\delta=0.015$ { to the curves} to avoid overlapping error bars.
We { plot $(1+\delta)$ in logarithmic units}
both to ease the comparison with similar plots in the literature and to
highlight deviations from linearity in the low-density regions. 
Error bars  represent the 2-$\sigma$ random scatter computed from the {\it Realistic} mocks.

The magnitude range that we are able to explore is set by competing constraints: 
the faint limit reflects the requirement of maximising the completeness of the sample 
whereas the bright limit is set by requiring{ $\langle N_g \rangle > 0.3 $ per cell.}
As a result,  the magnitude range shrinks with the redshift:
at $z=[0.5,0.7]$ it spans a range  $\Delta M_{\rm B} = 1.4$ 
whereas at  $z=[0.9,1.1]$  $\Delta M_{\rm B} =  0.5$.

In the upper plot the curves corresponding to the different magnitude cuts are well separated
for $\delta_g<0$. The {separation reduces and then disappears  with the} redshift.
{ This is not surprising since at $z \ge 0.9$ the luminosity range is very narrow, as we have seen.}
No significant trend with luminosity is seen at large over-density.
These { features, {or the lack of them,}} are robust to variations in the size of the cells in the range $R=[4,8]$ \hmpc  (see Table~\ref{tab:errors}) and confirm the {results obtained at lower redshifts from galaxy clustering}
(e.g. \citealt{norberg02,zehavi05,pollo06,coil08,skibba13,alhambra13,marulli13}),
 gravitational lensing (e.g. \cite{coupon12}) and  counts in cells (e.g. M05 and K11).

To further investigate { galaxy bias in} under-dense regions, we zoom into the $\delta<0$ range in Figure~\ref{fig:bias_zoom}.
The curves are the same as in Figure~\ref{fig:bias_mag}. The black long-dashed line represents the linear biasing function with 
a slope matching the $\hat{b}$ value estimated at $\delta_{MAX}$, which is listed in Table~\ref{tab:errors}.
Since $\hat{b}$ { only} weakly depends on the magnitude cut we only consider one representative case 
per panel.  The {local slope of the} biasing function is always steeper than the best-fitting linear bias model.
The horizontal, short-dashed line shows the  $\delta_g=-0.9$ threshold.The mass over-density at which {this line} crosses the biasing curves, $\delta_{\rm TH}$,  increases 
with the redshift and, to a lesser extent, with the luminosity.
This trend, which was noticed by M05 and, with less significance, by K11,has been interpreted as evidence that low-density regions are preferentially populated 
by low-luminosity galaxies. { Also, the quantity} $\delta_{\rm TH}$ has { been regarded}
as the typical mass over-density below which very few galaxies form. 

Figure~\ref{fig:bias_zoom} shows that galaxies can be found at mass over-densities well below $\delta_{\rm TH}$.
This low-density tail, together with the {steepness of the} biasing function for $\delta>\delta_{\rm TH}$,  
shows that the biasing relation in the under-density region significantly deviates from the linear prescription.
Non-linearity increases when decreasing the cell size. As we checked, for  $R=$ 4 \hmpc
the slope of the biasing curves further increases well above $\delta_{\rm TH}$. For $R=$ 8 \hmpc, { the difference 
disappears and the two slopes start to match. Still, } the {\it bias} curves keep featuring a negative $\delta$
 tail that cannot be matched by  linear models.


\begin{figure}
\includegraphics[width=0.5\textwidth]{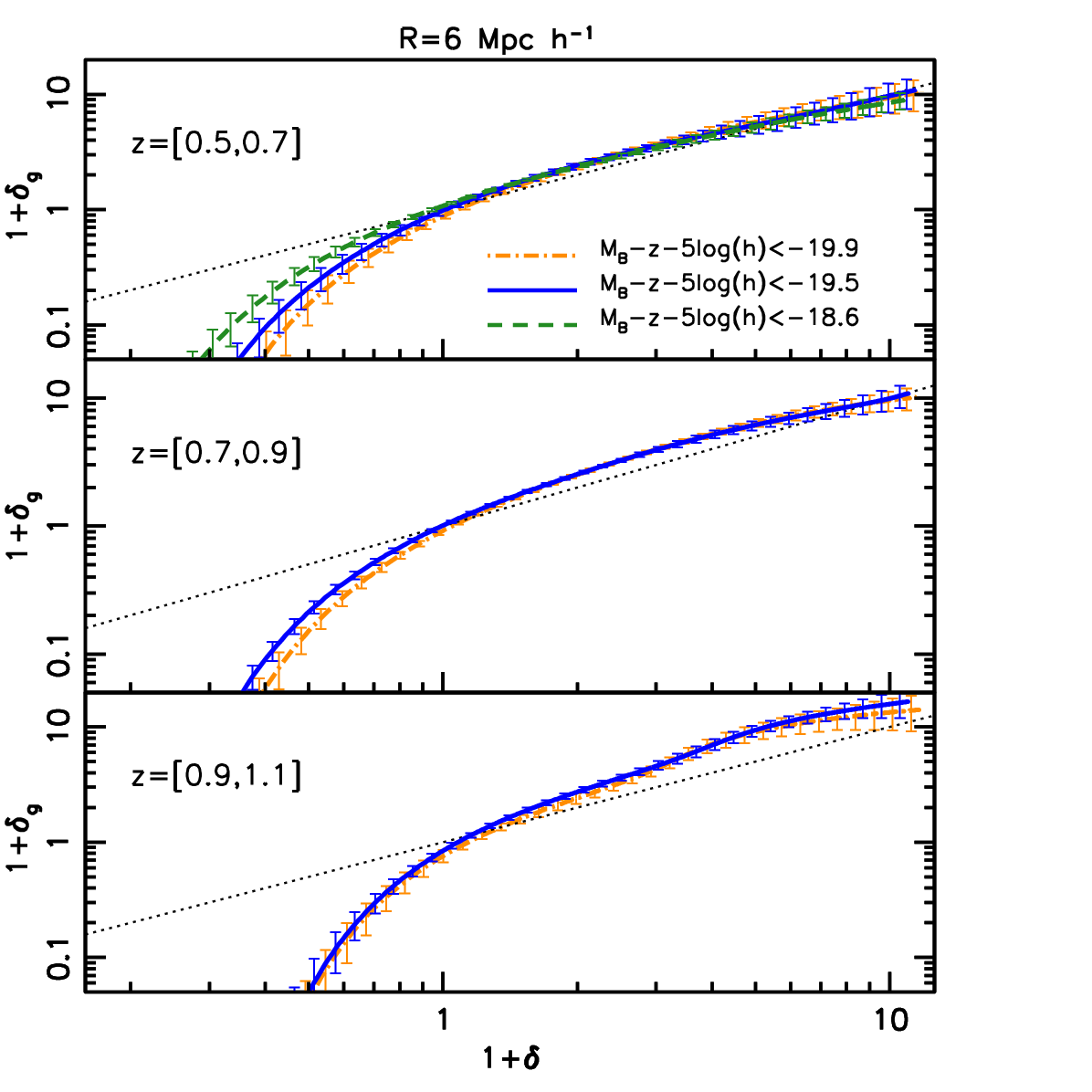}
\caption{
Mean biasing function of VIPERS galaxies from counts in cells of
radius  6  \Mpch  as a function of the B-band magnitude cut in three redshift ranges indicated in 
each panel.
Curves with different colours and line styles correspond to the different 
magnitude cuts indicated in upper panel. 
Error bars with matching colours represent the associated 1-$\sigma$ uncertainty intervals
estimated from the mocks.
{ A {horizontal }offset $\delta=0.015$ was applied to avoid overlapping error bars. 
{ All} biasing functions are plotted out to $\delta_{\rm MAX}$. }
}
\label{fig:bias_mag}
\end{figure}


\begin{figure}
\includegraphics[width=0.5\textwidth]{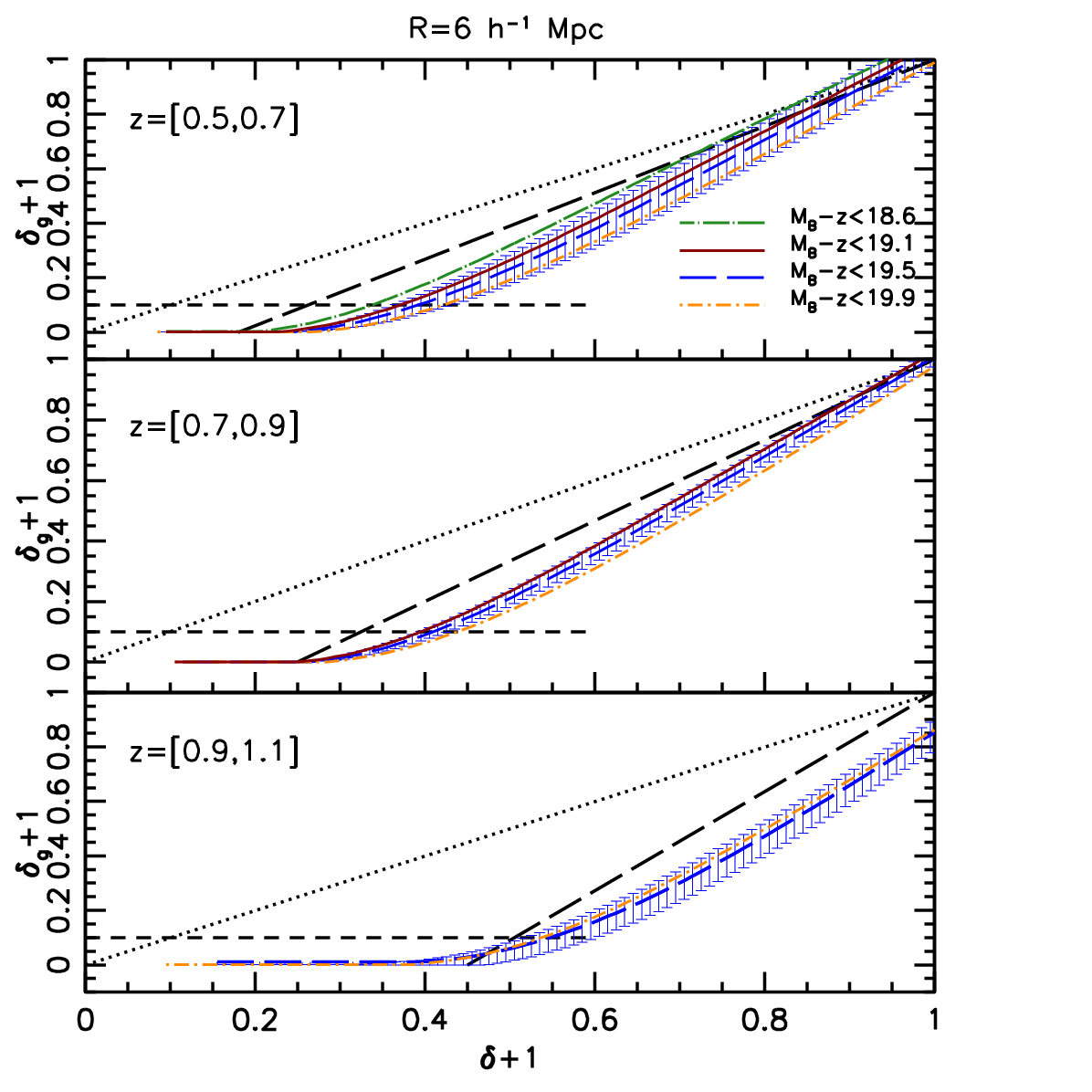}
\caption{
{ Zoom into the under-density range of  Fig.~\ref{fig:bias_mag}. The horizontal short-dashed line represents the 
over-density threshold $\delta_g=-0.9$. 
The long dashed line shows the linear biasing function 
$\delta_g=\hat{b}\delta$, for { the $\hat{b}$ value corresponding to the $M_{\rm B}-z-5log(h) < 19.5$ cut},
 which is listed in Table~\ref{tab:errors}.  The dotted line, shown for reference, shows the case { $b_{LIN}=1$}.  
Error bars represent the 1-$\sigma$ {\it rms }scatter among the mocks. They are only shown for the case
$M_{\rm B}-z-5log(h) < 19.5$ to avoid overcrowding.}
}
\label{fig:bias_zoom}
\end{figure}


Figure~\ref{fig:moments_mag} shows the second-order moment  $\hat{b}$ (left panels) and 
{ the ratio} $\tilde{b}/\hat{b}$ 
(right panels)  of the biasing functions shown in Fig.~\ref{fig:bias_mag}.
The same colour-code is used to indicate the magnitude cuts.
Large filled symbols refer to measurements performed at $\delta_{\rm MAX}$ 
assuming $\sigma_8=0.9$ whereas
the slightly offset, 
smaller open symbols refer to estimates performed at $\bar{\delta}_{\rm MAX}$.
The values of the corresponding bias moments are listed in Table~\ref{tab:errors}.
Error bars represent 1-$\sigma$ total random uncertainties estimated
from the {\it Realistic} mocks  (see Table~\ref{tab:errors}).

In the left panels of Fig.~\ref{fig:bias_mag}, 
we notice that{ in the low redshift bin, 
where the magnitude interval that we probe is larger, 
$\hat{b}$ increases with the luminosity.}
This {dependence} is much weaker {for} $z=[0.7,0.9]$ and completely absent 
at higher redshifts. 
We show results for cells of 6  \hmpc. { However,} the same trend is also seen for
4 and 8   \hmpc.

The right panels show the non-linear  parameter $\tilde{b}/\hat{b}$. Values that differ from unity
indicate deviations from linear bias (horizontal dashed line). A small but significant degree 
of non-linearity is present at all redshifts. { We} do not detect any significant 
dependence on luminosity { in any redshift bin and for any cell size.}


\begin{figure}
\includegraphics[width=0.5\textwidth]{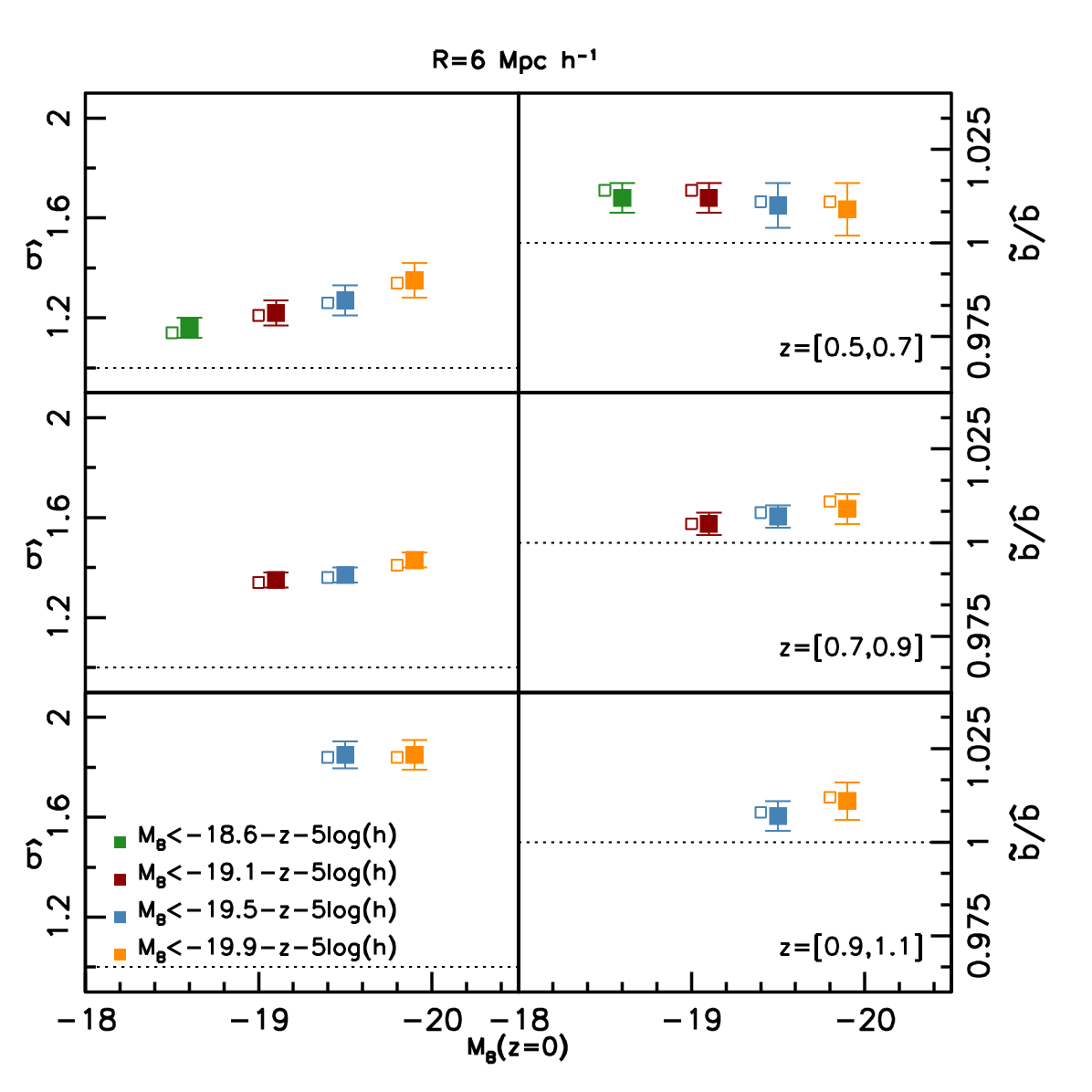}
\caption{
Second-order moments of the mean biasing functions shown in Fig.~\ref{fig:bias_mag}.
Left panels: {moment} $\hat{b}$. 
Right panels: non-linear parameter $\tilde{b}/\hat{b}$. The cell size is  $R=6$ \hmpc.
Error bars indicate  1-$\sigma$ scatter from the mocks.
The redshift ranges and colour code are the same { as} in Fig.~\ref{fig:bias_mag}. Magnitude cuts 
 are indicated in the plots.
All values were computed assuming $\sigma_8=0.9$.
The horizontal dashed line is plotted for reference and represents 
the case of no bias (left plot) and
linear bias (right panels).
{ Large filled symbols refer to measurements performed at $\delta_{\rm MAX}$,
and small, open symbols refer to estimates at $\bar{\delta}_{\rm MAX}$.}
}
\label{fig:moments_mag}
\end{figure}

{ A common feature of the reconstructed mean biasing functions at  $z=[0.9,1.1]$
is the presence of} some irregular behaviour (wiggling) at high over-densities.
This is the typical fingerprint of an imperfect inversion (Eq.~\ref{eq:inversions}) discussed in 
Section~\ref{sec:direct} and one of the reasons { for  introducing} the threshold 
$\delta_{\rm MAX}$. These irregularities typically arise as {a result of sampling 
rare, large over-densities with a limited number of independent cells.} 
The effect is most evident at large redshifts and for bright magnitude cuts,
i.e. when  the  sampling is sparser.
{ This affects the shape of the reconstructed mean biasing function. However, the impact
on the second moments}
 $\hat{b}$ and $\tilde{b}$ and, especially, $\tilde{b}/\hat{b}$, is {rather} limited.
This is because bias moments are integral quantities (Equation~\ref{eq:bhatdelta})
weighted by the mass PDF, which peaks at $\delta \sim 0$ and 
rapidly approaches zero in the high- and low-density tails.
Systematic errors in the bias reconstruction at large over-densities are {therefore} suppressed
when computing  $\hat{b}$ and $\tilde{b}$ and further smoothed out when computing their ratio.
 
Figure~\ref{fig:moments_vipers_mrange} demonstrates the validity of this conjecture.
In the left panels we  show the values of  
$\hat{b}(\delta)$ computed from equation~\ref{eq:bhatdelta}.
Curves with different line styles refer to the different magnitude cuts indicated in the plot.
Error bars with matching colours indicate the 1-$\sigma$ scatter from the mocks.
In the interval $z=[0.9,1.1]$ and for the brightest {and} sparsest sample,
$\hat{b}(\delta)$ flattens for  $\delta > 3$, { i.e.} well below  $\delta_{\rm MAX}$.
Analogous considerations hold for the curve $\tilde{b}/\hat{b}(\delta)$
shown in the right panels. These trends are robust to the size of the cells.
 }

\begin{figure}
\includegraphics[width=0.5\textwidth]{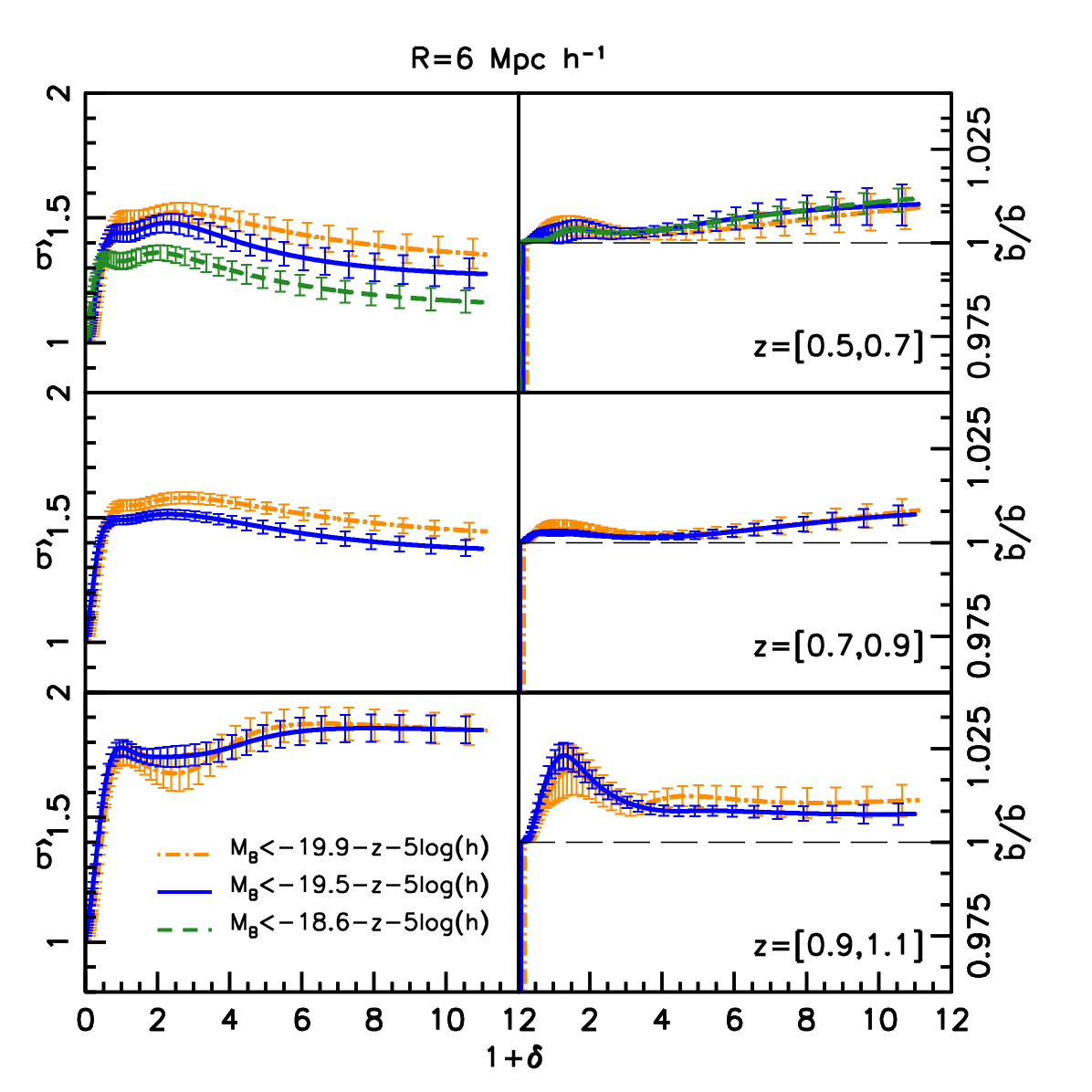}
\caption{
Left: second-order moment $\hat{b}(\delta)$ of the reconstructed  mean biasing functions shown in
Fig~\ref{fig:bias_mag}. The cell size is  $R=6$ \hmpc.
Different  line styles and colours indicate different luminosity cuts listed in the plot.  
The redshift ranges and  colour codes are the same as in  Fig~\ref{fig:bias_mag}.
Error bars represent  the {\it 1}-$\sigma$  scatter among the mocks.
Right: similar plots showing the non-linear bias parameter
 $\tilde{b}/\hat{b}(\delta)$.
{ A { horizontal} offset $\delta=0.015$ was applied to avoid overlapping error bars.
{ All } curves are plotted out to $\delta_{\rm MAX}$.}
}
\label{fig:moments_vipers_mrange}
\end{figure}


\subsection{Redshift dependence}
\label{sec:reds}

\begin{figure}
\includegraphics[width=0.5\textwidth]{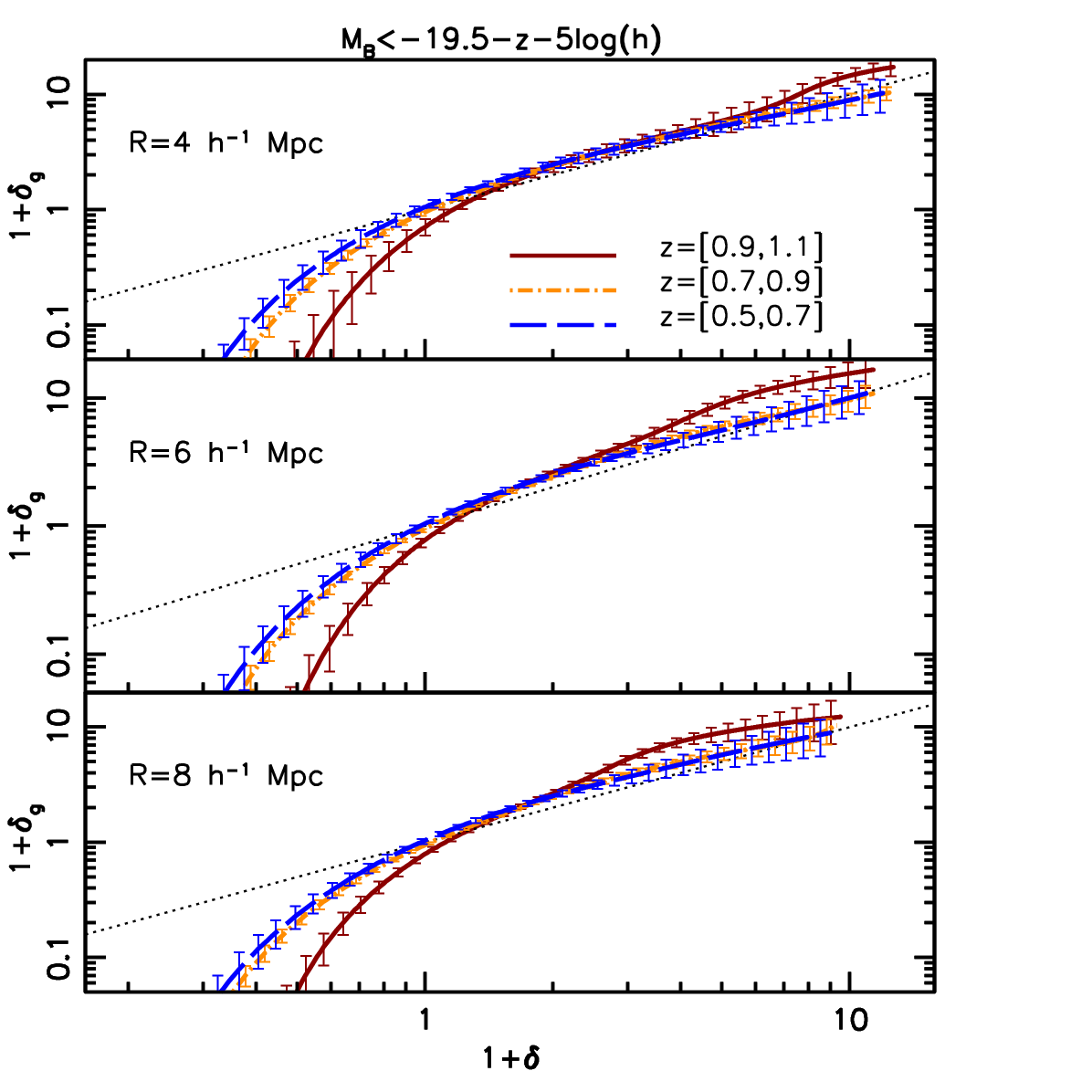}
\caption{
Mean biasing function of VIPERS galaxies with M$_{\rm B} <-19.5 -z -5\log (h)$
measured in different redshift bins, characterised by different colours and line-styles, as
indicated in the plot.
Error bars represents the 1-$\sigma$ scatter in the mocks.
The three panels refer to different cell sizes with radii $R=4, 6, 8$ \hmpc from top to bottom.
{ A horizontal  offset $\delta=0.015$ was applied to avoid overlapping error bars.
All curves are plotted out to $\delta_{\rm MAX}$.}
}
\label{fig:bias_reds}
\end{figure}

To explore the bias dependence on the redshift we set a magnitude cut
M$_{\rm B} =-19.5 -z -5\log (h)$ and estimated the mean biasing function of brighter galaxies
in the three redshift bins. This $z$-dependent magnitude cut is designed to 
account for luminosity evolution \citep{zucca09}, so that differences in the galaxy bias
measured in the different $z$-bins can be interpreted as the result of a genuine evolution.
The results of our analysis are shown in Figure~\ref{fig:bias_reds}.
The plots are analogous to those of Fig.~\ref{fig:bias_mag} and use the same symbols, colour scheme, and line style.
However, we consider cells of different sizes in the three panels.

The biasing function shows little or no evolution in the range $z=[0.5,0.9]$, as demonstrated by the 
proximity between the dashed-blue ($z=[0.5,0.7]$) and dot-dashed orange ($z=[0.7,0.9]$) curves and 
the overlap of their 1-$\sigma$ error bars.
 The red solid line, however, is separated fully
  from the others, indicating that
 galaxy bias evolves significantly beyond  $z=0.9$.
 This evolution is detected both in low- and high-density environments.
 { { It} implies that $\delta_{\rm TH}$  increases significantly with the redshift, 
 indicating that evolution shifts galaxy formation towards regions of progressively lower density. 
 }
 At  $\delta > 0$ the effect of evolution is that of increasing the slope of the biasing function with $z$.
 Since in this range the biasing is close to linear, {an estimate of $b_{LIN}$ would reveal a redshift evolution
 consistent with that observed in several analyses}, as detailed in Section~\ref{sec:comparison}.
 The same trend is evident in all panels,
 indicating that the bias evolution is similar in all explored scales.

{ At high redshifts and for $R=8$ \hmpc (bottom panel) 
the biasing function}
is characterised { by some irregularities}  at moderate values of $\delta$.
{ As pointed out, these  have 
little impact on the estimated} values of $\hat{b}$ and  $\tilde{b}/\hat{b}$. 
 It is reassuring that these anomalies are only seen at high redshifts, confirming the
fact that they are induced by poor sampling of the counts probability.
{ All this} makes us confident that bias evolution is a genuine feature.


\begin{figure}
\includegraphics[width=0.5\textwidth]{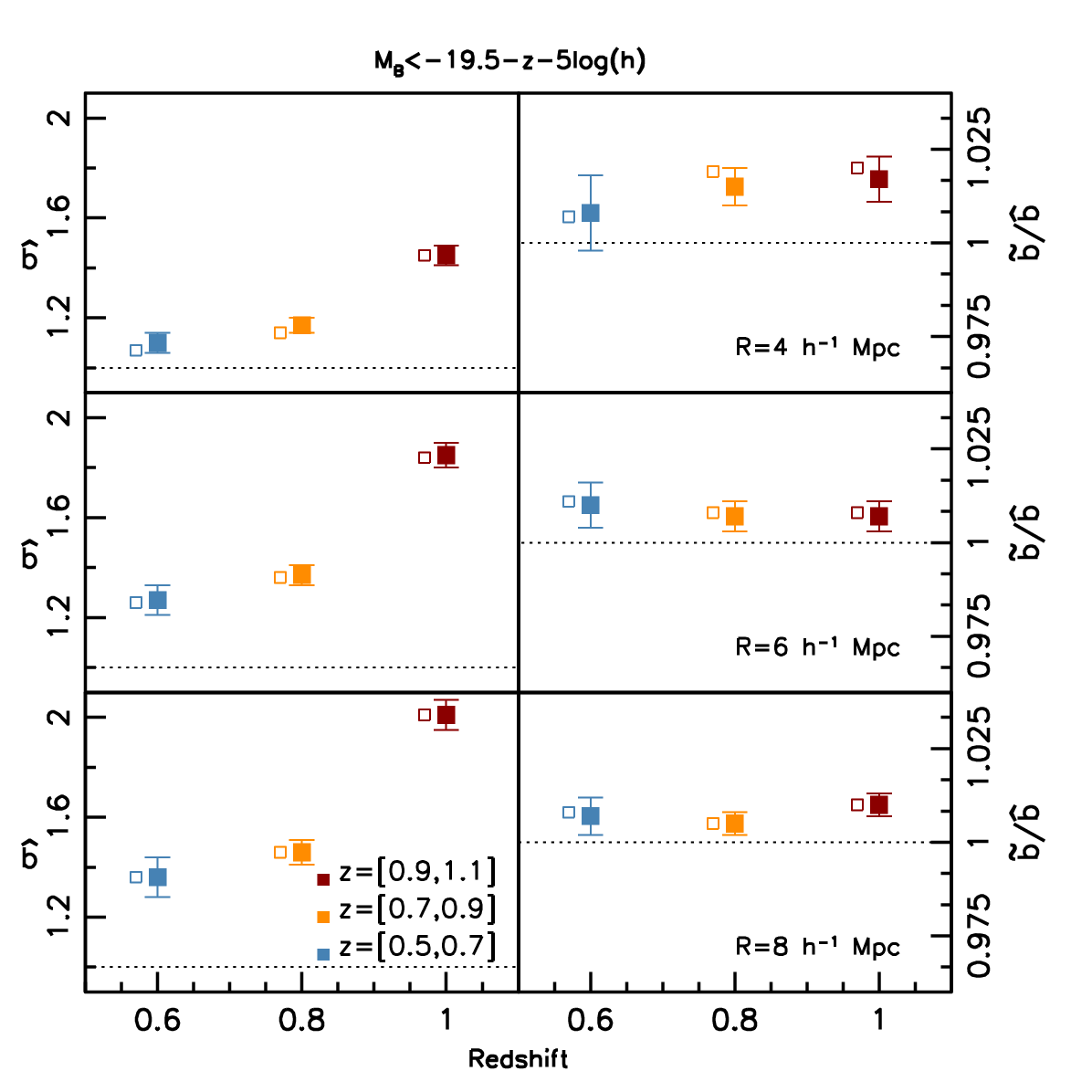}
\caption{
Moments of the mean biasing functions shown in Fig.~\ref{fig:bias_reds}.
Panels from top to bottom indicate cells of increasing size.
Left panels: $\hat{b}$. 
Right panels: $\tilde{b}/\hat{b}$. 
Error bars indicate  the 1-$\sigma$ scatter from the mocks.
All values were computed assuming $\sigma_8=0.9$.
{ Large filled symbols refer to measurements performed at
  $\delta_{\rm MAX}$, small open symbols refer to estimates at $\bar{\delta}_{\rm MAX}$.}
}
\label{fig:moments_reds}
\end{figure}

{ Figure~\ref{fig:moments_reds} shows the values of $\hat{b}$ and  $\tilde{b}/\hat{b}$ 
as a function of  redshift. The values were obtained by integrating the 
mean biasing functions in Figure~\ref{fig:bias_reds} out to the values $\delta_{\rm MAX}$ and
$\bar{\delta}_{\rm MAX}$  { (large and small symbols, respectively).
The corresponding values are listed in  Table~\ref{tab:errors}.}}
The colour code is the same as in Figure~\ref{fig:bias_reds} and is indicated in the plot.
The mean slope of the curve, $\hat{b}$ (left panels),  increases significantly beyond $z=0.9$ whereas we see little or no evolution 
at lower redshifts. This {shows that the} trend seen in Figure~\ref{fig:bias_reds} {is seen at}
all scales, indicating that the bias evolution at $z>0.9$ is indeed a robust feature.
The bias parameter of VIPERS galaxies brighter than M$_{\rm B}= -19.9 -z -5\log (h)$ 
exhibits a small but significant degree of non-linearity  at all redshifts and  scales explored in our analysis
(right panels); this bias parameter{, however, does not { significantly} evolve with redshift}.
 These results are robust to the luminosity cut since they are also found for galaxies brighter than  M$_{\rm B} <-19.9 -z -5\log (h)$.

\subsection{Scale dependence}
\label{sec:scale}

\begin{figure}
\includegraphics[width=0.5\textwidth]{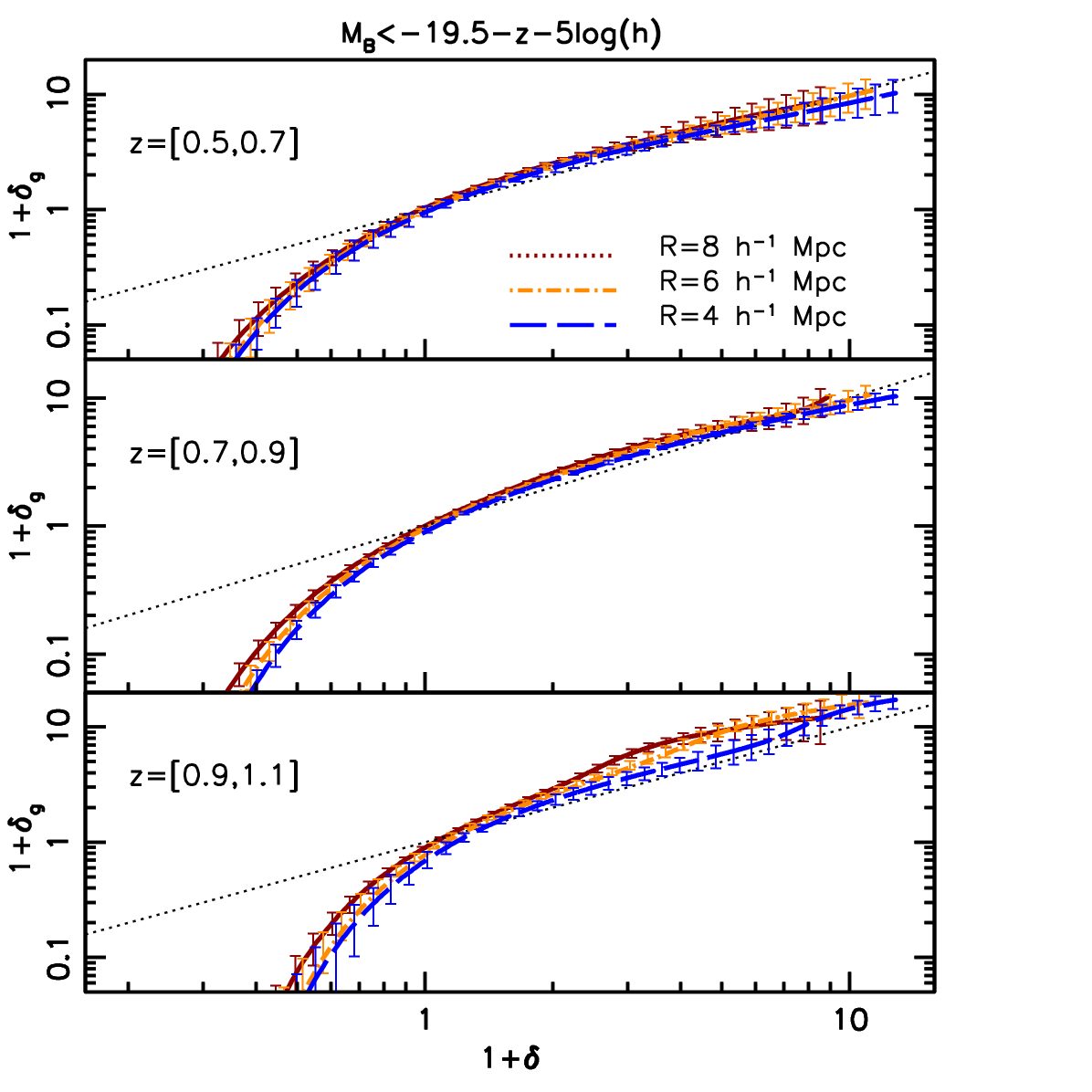}
\caption{
Mean biasing function of VIPERS galaxies with M$_{\rm B} <-19.5 -z -5\log (h)$
computed from counts in cells with radii of $4, 6 \, {\rm ,and} \, 8$ \hmpc. 
Biasing functions at different scales are indicated with different colours and line styles, as indicated in the plots.
Error bars represent the 2-$\sigma$ {\it rms} scatter in the mocks.
Different panels refer to different redshift shells.
{ An offset $\delta=0.015$ was applied to avoid overlapping error bars.
Curves are plotted out to $\delta_{\rm MAX}$.}
}
\label{fig:bias_scale}
\end{figure}

In Fig.~\ref{fig:bias_scale} we explore the dependence of the bias of VIPERS galaxies on the radius of the 
cells down to a scale of 4 \hmpc.
In the plots we show the mean biasing function of VIPERS galaxies brighter than M$_{\rm B} =-19.5 -z -5\log (h)$
measured at $R=4, 6 \, {\rm and} \, 8$ \hmpc. Different scales are characterised by different colours, as
indicated in the plot. The panels show the results in the three redshift shells.
{ At negative over-density the curves are remarkably similar, indicating that 
 $\delta_{\rm TH}$ and the efficiency of galaxy formation do not depend on the scale in the range $[4,8]$ \hmpc.
 At  $\delta >0$ the curves steepen with the radius of the cell, indicating that biasing increases with the scale
  especially at high redshift.}

\begin{figure}
\includegraphics[width=0.5\textwidth]{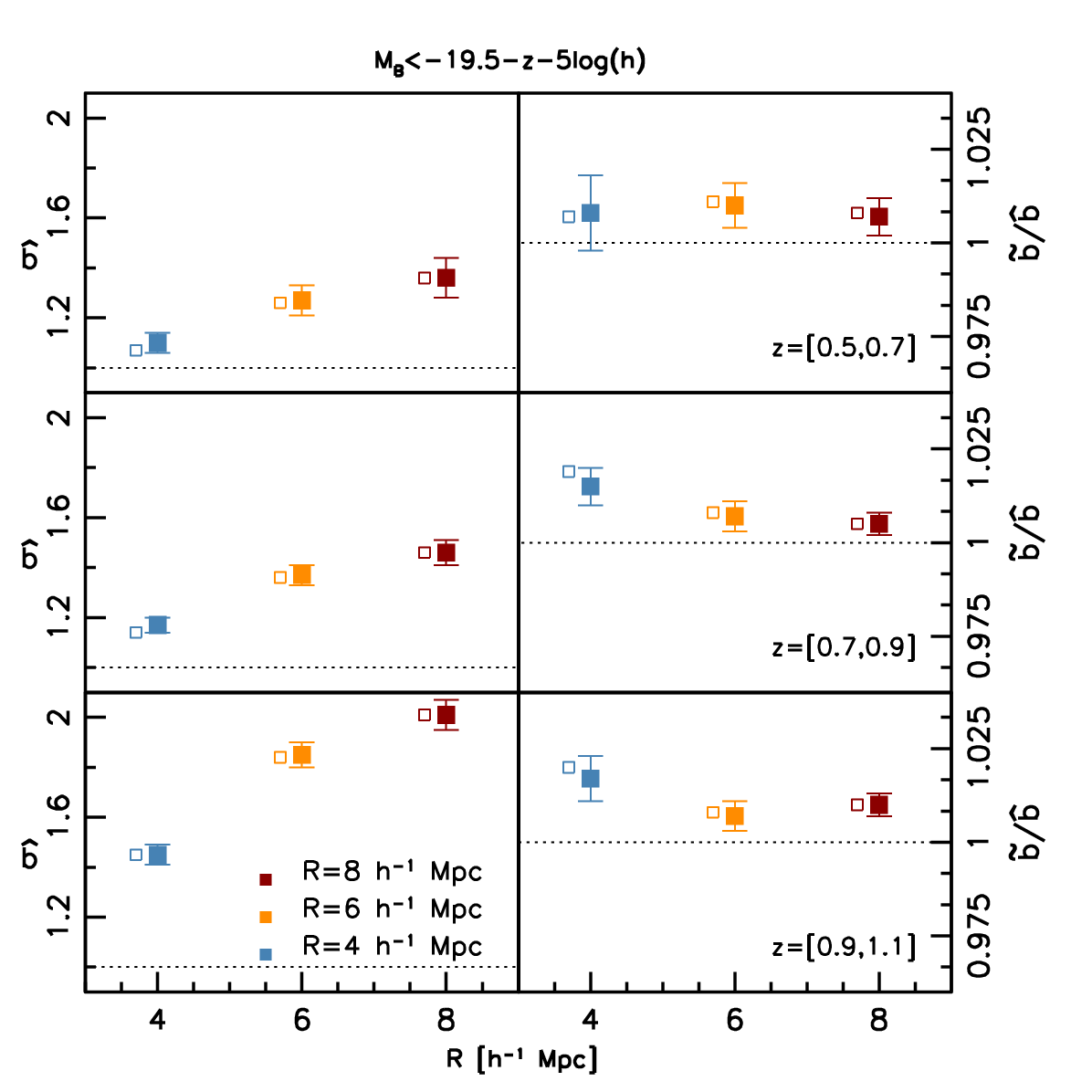}
\caption{
Moments of the mean biasing functions
vs. the size of the cell.
Panels from top to bottom refer to different redshift ranges indicated in the plot.
Left panels: $\hat{b}$. 
Right panels: $\tilde{b}/\hat{b.}$ 
The parameters were computed assuming $\sigma_8=0.9$.
Error bars indicate  the 1-$\sigma$ {\it rms} scatter from the mocks.
{ Large, filled symbols refer to measurements performed at $\delta_{\rm MAX}$,
small open symbols refer to estimates at $\bar{\delta}_{\rm MAX}$.}
}
\label{fig:moments_scale}
\end{figure}

A more quantitative assessment of scale dependence is shown in Fig.~\ref{fig:moments_scale}. 
The value of $\hat{b}$ steadily increases with the cell radius, $R$, especially 
at high redshift.
This trend may sound counterintuitive; galaxies are expected to trace 
the mass with increasing accuracy on a larger scale and, consequently,
galaxy bias is expected to { approach its linear value}.
This, however, occurs on scales much larger than those
considered here (see e.g. \citealt{Wild}).
On the scales explored here the halo model predicts that the 
opposite trend should be observed (see e.g.  Fig. 4 of  \citealt{zehavi04}).
The reason is that in this range of scales the  contribution to galaxy clustering of the 1-halo term, which dominates on small 
scales, is comparable to that of the 2-halo term, which dominates on large scales.
The scale of the crossover depends on galaxy type and redshift but it is expected 
to be bracketed in the range probed by our analysis.
This explanation is corroborated by the fact that the values of $\hat{b}$ measured in 
the HOD mocks, designed following the halo model prescriptions, do
show an increasing trend with the size of the cells.

An increase of galaxy bias with the scale was already detected at lower redshifts 
from the analysis of galaxy clustering  \citep{zehavi05} and from weak lensing 
\citep{hoekstra02,simon07}. This is the first detection
at relatively high redshift that exploits counts in cell statistics.
A small, but significant amount of non-linearity is detected at all redshifts. 
Unlike $\hat{b}$, the non-linear parameter $\tilde{b}/\hat{b}$ seems to be scale independent.
These results are robust to magnitude cut since similar trends for $\hat{b}$ and  $\tilde{b}/\hat{b}$ are also seen when one
restricts the biasing analysis to objects brighter than M$_{\rm B} <-19.9 -z -5\log (h)$.

\subsection{Results from the whole dataset}
\label{sec:full}

\begin{figure}
\includegraphics[width=0.52 \textwidth]{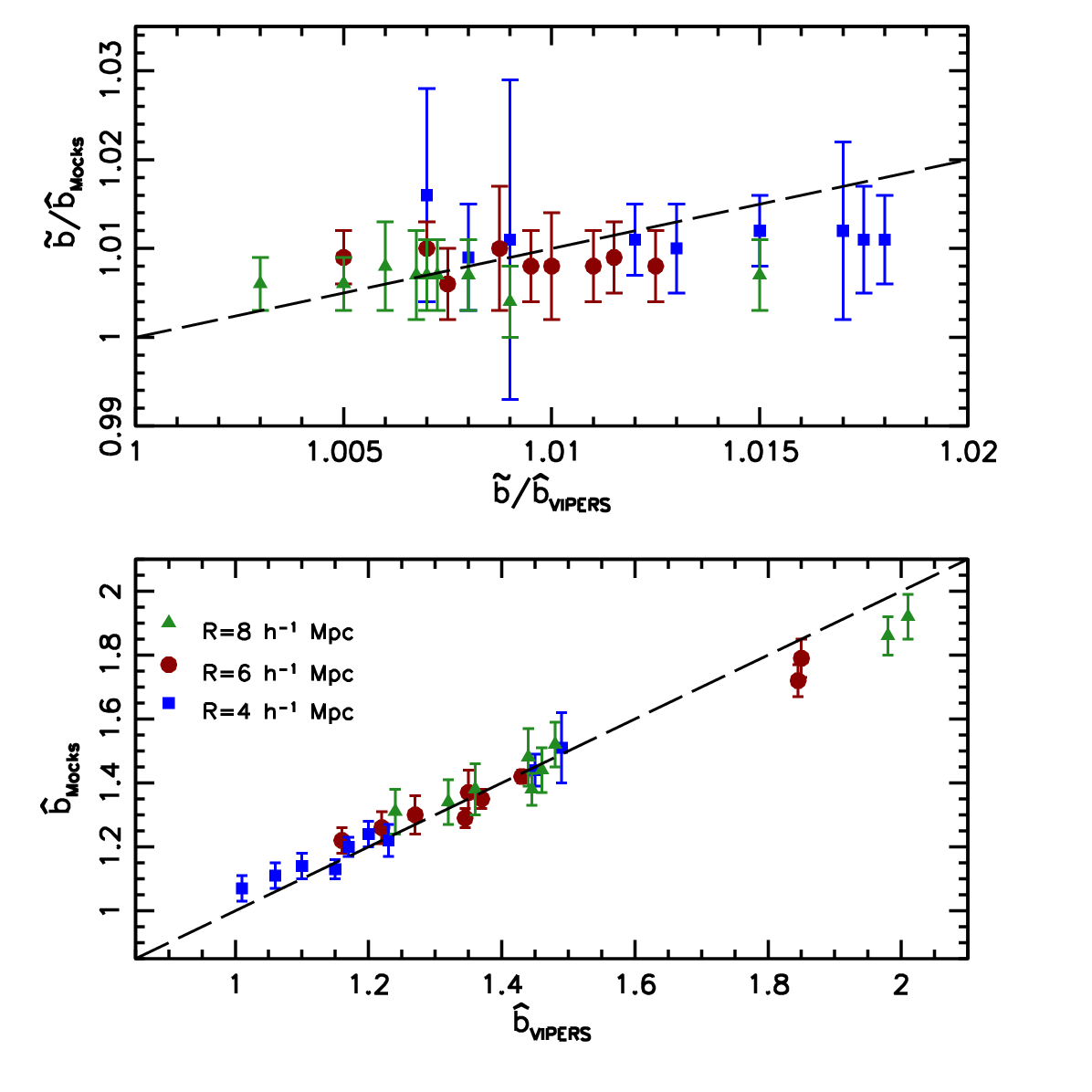}
\caption{ 
{  Bottom panel:  comparison between $\hat{b}$  measured in various mock subsamples  and $\hat{b}$ 
 measured in the} VIPERS catalogue. Different symbols and colours refer to results 
obtained with cells of different sizes, as indicated in the plot. The subsamples were obtained
by applying the same magnitude and redshift cuts used in this section and indicated in Table~\ref{tab:errors}.   
The error bars represent 1-$\sigma$  scatter from the mocks.
Top panel: comparison among the non-linear parameters  $\tilde{b}/\hat{b}$ measured in the mocks and in the real sub-catalogues}.
\label{fig:relamock}
\end{figure}

{ We now summarise the results {presented in this section.
Overall, the biasing functions of the VIPERS subsample
are in qualitative agreement with those of M05 and K11 with some 
intriguing differences.
At moderate over-densities and out to $\delta_{\rm MAX}$, 
our  biasing functions are close to linear
with a slope close to $\hat{b}(\delta_{\rm MAX})$. This is at variance with 
M05 and K11 whose biasing function  flattens at large $\delta$, leading to 
an anti-bias signature.
This feature has been variously interpreted as evidence for 
quenching processes  \citep{blanton00},  enhanced galaxy merging rate  \citep{marinoni05},
and early galaxy formation  \citep{yosh01} in high-density regions. 
We find a similar flattening only if we push our analysis 
beyond  $\delta_{\rm MAX}$. However, its statistical significance is 
less that 2 $\sigma$. A similar feature was also
detected in simulations and interpreted as an artefact due to limitations of the bias 
estimator at high redshift \citep{Sigad00}.
Given the fact that all these works, including ours, use a similar technique to 
measure galaxy bias, we  suspect that the flattening at large density is not a genuine effect. }

At $\delta <0$ the biasing function significantly deviates from linear prescription and is steeper than 
$\hat{b}$. Moreover, the galaxy density remains positive below $\delta_{\rm TH}$, indicating that galaxy formation 
is not entirely quenched { even}  in { very}  low-density environments.}

Table~\ref{tab:errors} lists the bias parameters measured in the VIPERS subsamples of Table~\ref{tab:cats}
together with random and systematic errors estimated from the mocks.
We computed all of the parameters by integrating the mean biasing 
function out to the value  $\delta_{\rm MAX}$ listed in the Table. 
Altogether the results confirm the various trends that we described in the  previous sections:
the value of  $\hat{b}$ increases with luminosity,  scale,  and with the redshift  beyond $z = 0.9$.
Deviations from linear biasing are small but typically detected with significance larger than 1 $\sigma$.
The non-linear bias parameter is, within the errors, independent of redshift, luminosity, or scale.
 
{ We obtained errors from {the} VIPERS mock catalogues  designed to match
the 2-point statistics of real galaxies but not their abundance or their bias.
{ One consequence of this is that bright galaxies in the subsamples of the mocks are sparser than the real galaxies at $z<0.9$
 (see Table~\ref{tab:cats}).}    
As a result our random errors somewhat overestimate the real errors
.
{ As for the bias, if that of the galaxies in the real sample is different from the mock sample, then our error estimate 
would be affected.}
 We compared the values of $\hat{b}$ and $\tilde{b}/\hat{b}$ in the mock and in the real samples to investigate this issue.
The resulting scatter plots are shown in  Figure~\ref{fig:relamock}.
The different points represent the individual subsamples considered in our analysis. Symbols with different colours 
are used to highlight results obtained with different cell sizes.
Error bars represent  the {\it rms} scatter in the mocks. 
Most of the points deviates less than 2 $\sigma$ from the expected value (black dashed line), 
implying that our mocks are realistic and that our errors are indeed reliable.}

\begin{table*}
\caption{Bias parameters of VIPERS galaxies  and their errors.}     
\label{tab:errors}      
\centering          
\begin{tabular}{c c c c c c c c c c c c}     
\hline\hline    
$z$-range                  & M$_{\rm B}$- cut                       & R                                             & $\delta_{\rm MAX}$                 & 
$\hat{b} $                   & $\sigma_{\rm RND}^{\hat{b}} $ & $\sigma_{\rm CV}^{\hat{b}} $ & $\sigma_{\rm SYS}^{\hat{b}} $ &  
$\tilde{b}/\hat{b} $ & $\sigma_{\rm RND}^{NL} $            & $\sigma_{\rm CV}^{NL}   $          & $\sigma_{\rm SYS}^{NL} $           \\
                                    & M$_{\rm B}(z=0)-5\log h$                & \Mpch                                   &                                                &
                                  &                                                       &                                                     &                                                    &
                                  &                                                       &                                                     &                                                    \\                                  
\hline
0.5-0.7 & $-18.6-z$ &  4 & 11(15)&   1.01 (0.98) & 0.04 &  0.02 &  0.02  & 1.018 (1.021)&  0.005 &  0.003 &  -0.003 \\
0.5-0.7 & $-19.1-z$ &  4  & 11(15)& 1.06 (1.03)&  0.04 &  0.02 &  0.02  & 1.017 (1.021)& 0.006 &  0.002 &  -0.004 \\
0.5-0.7 & $-19.5-z$ &  4  &  11(15)& 1.10 (1.07)&  0.04 &  0.02 &  0.02 &  1.017 (1.020) &  0.010 &  0.002 &  -0.005 \\
0.5-0.7 & $-19.9-z$ &  4  &  11(15)& 1.23 (1.20)&  0.05 &  0.02 &  0.02 &  1.007 (1.009) &  0.012 &  0.002 &  -0.006 \\
0.7-0.9 & $-19.1-z$ &  4 &  11(15)& 1.15 (1.12)& 0.03 & 0.02 & 0.01 & 1.012 (1.018)& 0.004 & 0.002 & -0.002 \\
0.7-0.9 & $-19.5-z$ &  4  &  11(15)& 1.17 (1.14)& 0.03  & 0.02 & 0.02 & 1.015 (1.019)& 0.004 & 0.001 & -0.002 \\ 
0.7-0.9 & $-19.9-z$ &  4  &   11(15)& 1.20 (1.15) & 0.04  & 0.02 & 0.02 & 1.013 (1.018) & 0.005 & 0.001 & -0.002 \\
0.9-1.1 & $-19.5-z$ &  4  &  11(15)&  1.45 (1.45)  & 0.05  & 0.02  & 0.02 & 1.008 (1.007) & 0.006 & 0.001 & $-0.002$ \\
0.9-1.1 & $-19.9-z$ &  4  &  11(15)& 1.49 (1.46) & 0.11 & 0.02 & 0.03 & 1.008 (1.008) & 0.018 & 0.001 & $-0.003$ \\
\hline
0.5-0.7 & $-18.6-z$ &  6 & 10(14)& 1.16 (1.14) & 0.04 & 0.04 & 0.03 & 1.012 (1.014) & 0.004 & 0.002 & -0.003 \\
0.5-0.7 & $-19.1-z$ &  6  &  10(14)& 1.22 (1.21) & 0.05 & 0.04 & 0.03 & 1.012 (1.014) & 0.004 & 0.002 & -0.003 \\ 
0.5-0.7 & $-19.5-z$ &  6  &  10(14)& 1.27 (1.26) & 0.06 & 0.04 & 0.04 & 1.010 (1.011) & 0.006 & 0.002 & -0.004  \\ 
0.5-0.7 & $-19.9-z$ &  6  &  10(14)& 1.35 (1.34) & 0.07 & 0.05 & 0.05 & 1.009 (1.011) & 0.007 & 0.002 & -0.004  \\ 
0.7-0.9 & $-19.1-z$ &  6 &   10(14)& 1.35 (1.34) & 0.03 & 0.03 & 0.02 & 1.005 (1.005) & 0.003 & 0.002 & -0.002 \\ 
0.7-0.9 & $-19.5-z$ &  6  &  10(14)& 1.37 (1.36) & 0.03 & 0.02 & 0.02 & 1.007 (1.008) & 0.003 & 0.002 & -0.002 \\
0.7-0.9 & $-19.9-z$ &  6  &  10(14)& 1.43 (1.41) & 0.03 & 0.03 & 0.02 & 1.009 (1.011) & 0.004 & 0.002 & -0.002 \\
0.9-1.1 & $-19.5-z$ &  6  &   10(14)& 1.85 (1.84) & 0.05 & 0.03 & 0.01 & 1.007 (1.008) & 0.004 & 0.001 & -0.002 \\
0.9-1.1 & $-19.9-z$ &  6  &   10(14)& 1.85 (1.84) & 0.06 & 0.03 & 0.02 & 1.011 (1.012) & 0.004 & 0.001 & -0.002 \\
\hline
0.5-0.7 & $-18.6-z$ &  8 & 8(12)&1.24 (1.23)  & 0.07 & 0.05 & 0.04 & 1.007 (1.008) & 0.005 & 0.003  & -0.003 \\
0.5-0.7 & $-19.1-z$ &  8  &  8(12)&  1.32 (1.31) & 0.07 & 0.05 & 0.04 & 1.007 (1.008) & 0.004 & 0.002  & -0.003 \\
0.5-0.7 & $-19.5-z$ &  8  &  8(12)&  1.36 (1.36)  & 0.08 &  0.06 & 0.05 & 1.007 (1.008) & 0.005 & 0.002 &  -0.003 \\
0.5-0.7 & $-19.7-z$ &  8  &  8(12)&  1.40 (1.39)  & 0.09 &  0.06 & 0.06 & 1.006 (1.007) & 0.005 & 0.002 &  -0.003 \\
0.5-0.7 & $-19.9-z$ &  8  &  8(12)&  1.44 (1.44)  & 0.09 &  0.06 & 0.06 & 1.006 (1.006) & 0.005 & 0.002 &  -0.003 \\
0.7-0.9 & $-19.1-z$ &  8 &  8(12)& 1.44 (1.44) & 0.04 & 0.04 & 0.02 & 1.003 (1.003) &  0.003 & 0.002 & -0.002 \\
0.7-0.9 & $-19.5-z$ &  8  & 8(12)&  1.46 (1.46) & 0.05 & 0.04 & 0.03 & 1.005 (1.005) & 0.003 & 0.002 & -0.002 \\
0.7-0.9 & $-19.7-z$ &  8  & 8(12)&  1.48 (1.48) & 0.07 & 0.04 & 0.03 & 1.008 (1.008) & 0.004 & 0.002 & -0.002 \\
0.7-0.9 & $-19.9-z$ &  8  & 8(12)& 1.51 (1.50) & 0.08 & 0.04 & 0.04 & 1.008 (1.009) & 0.006 & 0.002 & -0.002 \\  
0.9-1.1 & $-19.5-z$ &  8  & 8(12)& 2.01 (2.01) & 0.06 & 0.04 &  0.02 & 1.010 (1.010) & 0.003 & 0.002 & -0.002 \\
0.9-1.1 & $-19.7-z$ &  8  & 8(12)& 1.98 (1.98) & 0.06 &  0.04 & 0.03 & 1.009 (1.009) &  0.003 & 0.003 & -0.002 \\
0.9-1.1 & $-19.9-z$ &  8  & 8(12)& 2.01 (2.01) & 0.07 &  0.05 & 0.04 & 1.015 (1.015) &  0.004 & 0.003 & -0.001 \\
0.7-0.9 & $-20.0$ &  8 & 8(12)& 1.43 (1.42) & 0.04 & 0.03 & 0.02 & 1.006 (1.005) &  0.002 & 0.002  & -0.001 \\
\hline \hline
\end{tabular}
\tablefoot{ Col. 1: redshift range. Col. 2: $z$-dependent B-band magnitude cut. 
Col. 3: cell radius [\hmpc].
Col.  4: maximum over-density considered in the analysis $\delta_{\rm MAX}$; the value in parenthesis indicates  $\bar{\delta}_{\rm MAX}$.
Col. 5: estimated value of  the bias moment $\hat{b}$; the values in parenthesis refer to measurements performed at $\bar{\delta}_{\rm MAX}$. 
Col. 6: total random error on $\hat{b}$. 
Col. 7: cosmic variance contribution to $\hat{b}$ error.  Col. 8: systematic error on $\hat{b}$.
Col. 9: estimated value of  the non-linearity parameter $\tilde{b}/\hat{b}$; values in parenthesis refer to measurements performed at $\bar{\delta}_{\rm MAX}$.
Col. 10: total random error on $\tilde{b}/\hat{b}$.
Col. 11: cosmic variance contribution to $\tilde{b}/\hat{b}$ error.  Col. 12: systematic error on $\tilde{b}/\hat{b}$.
}
\end{table*}

\section{Comparison with previous results}
\label{sec:comparison}

Several authors estimated the bias of galaxies in the same range, $z=[0.5,1.1]$, considered here. The majority of these authors assumed linear bias and estimated the bias parameter from galaxy clustering 
\citep{coil06,meneux06,coil08, meneux08,meneux09,coupon12,marulli13,skibba13,alhambra13}. Only
a handful of papers addressed the issue of non-linear  or scale-dependent bias at these redshifts
(M05, K11, \cite{simon07,jullo12}).
In this section, we compare our results with both types of analyses. First we compare our estimated non-linear bias parameter
with available measurements from previous studies.
Then we consider the most recents estimates of the linear bias parameter $b_{\rm LIN}$ in this redshift range 
available in the literature and compare them with our value of $\hat{b}$.In these comparisons all results were rescaled to the value $\sigma_8=0.9$ adopted in this paper
whenever required.

\subsection{Galaxy bias from counts in cells}
\label{sec:pdfcomparison}

\begin{figure}
\includegraphics[width=0.54\textwidth]{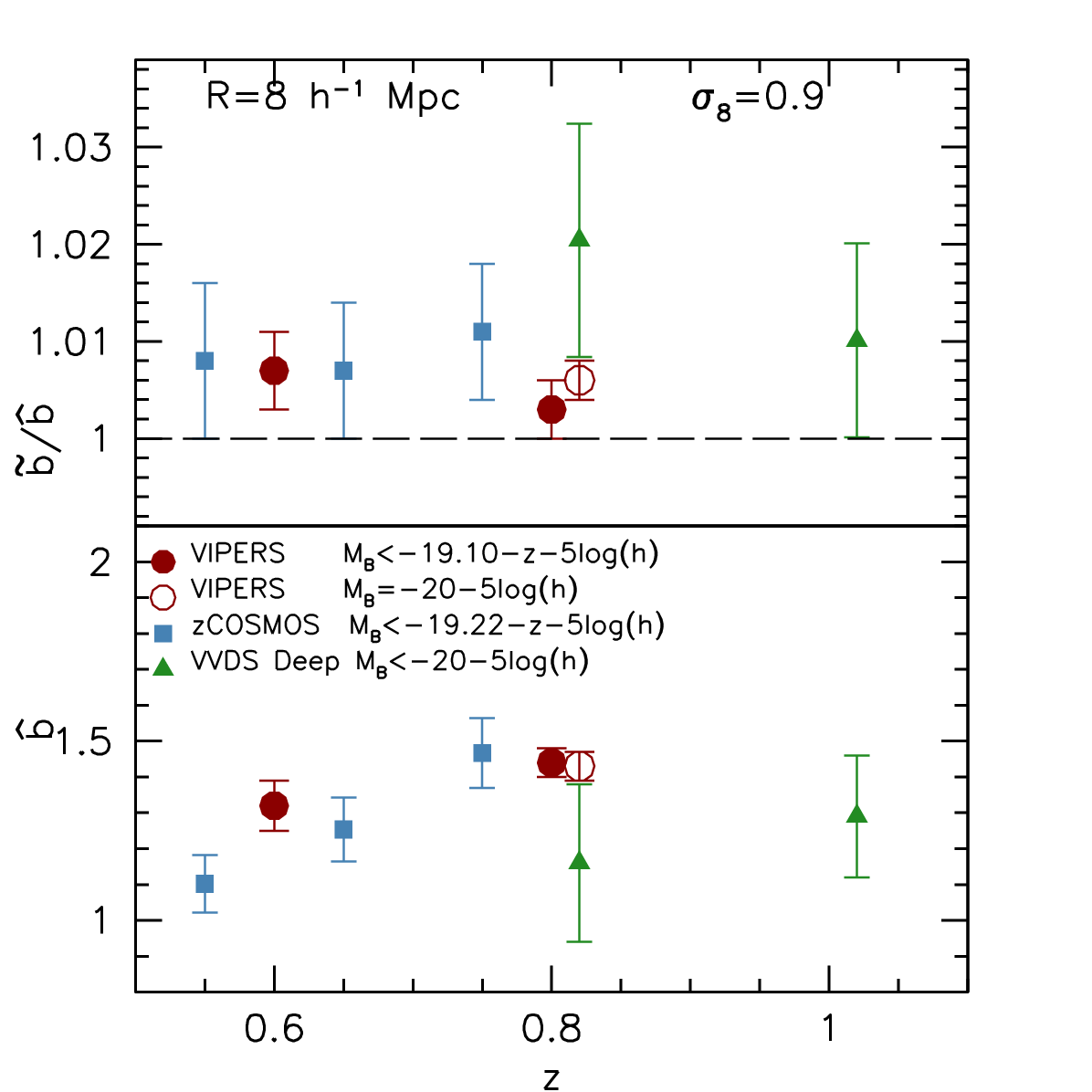}
\caption{
Comparison among  the values of $\tilde{b}/\hat{b}$  (top panel)
and 
  $\hat{b}$ (bottom panel)  for zCOSMOS galaxies (blue squares)
for VVDS-Deep galaxies (green triangles) and VIPERS  galaxies (filled red circles).
All samples are luminosity limited and the magnitude cuts are indicated in the plot.
The open red circles represents a VIPERS subsample matching the magnitude cut and 
redshift range of the VVDS-Deep sample.
Estimates for the bias parameters  of  zCOSMOS are taken from K11,
and those for VVDS-Deep galaxies are from M05.
}
\label{fig:bias_compf}
\end{figure}

In Figure~\ref{fig:bias_compf} we plot the values of $\hat{b}$ and $\tilde{b}/\hat{b}$ obtained from our analysis
as a function of redshift (filled { and open} red dots) and compare these values to those obtained by  M05 (green triangles)
and K11 (blue squares) from counts in cells following a procedure similar to ours.
We do not consider the results of the analyses of \cite{simon07} and \cite{jullo12} since these authors 
estimate the so-called correlation parameter that accounts for both non-linearity and stochasticity. 

We only considered objects that, at a given redshift, 
span a similar range of magnitudes to avoid mixing evolution and luminosity dependence. For VIPERS we consider objects with M$_{\rm B}<-19.1-z-5\log(h)$.
For zCOSMOS we consider objects above a similar cut-off, M$_{\rm B}=-19.22-z-5\log(h)$.
For the VVDS-Deep sample, M05 use a redshift-independent luminosity threshold  of M$_{\rm B}=-20.0-5\log(h),$ which is comparable 
with the above cut-offs in the range $z=[0.8,1.1]$.
We 
considered an additional VIPERS subsample cut at the same constant magnitude limit as M05 to improve the consistency in the comparison with VVDS.

In the case of M05, the values of  $\hat{b}$ and $\tilde{b}/\hat{b}$
shown in {Figure~\ref{fig:bias_compf}}
 were inferred from the published values 
of $\tilde{b}$ and  $\hat{b}/\tilde{b}$.
In addition, M05 do not provide the errors for $\tilde{b}/\hat{b}$.
The error bars shown in the plot  were extrapolated from the errors on $\hat{b}$ under the assumption 
that the ratio of the errors on $\hat{b}$ and those on $\tilde{b}/\hat{b}$ { are} the same for the two 
datasets. The comparison between zCOSMOS and VIPERS shows that this assumption is approximately 
valid.
The zCOSMOS points are plotted at the centre of their redshift bins. In the VVDS case we 
added an offset $\Delta_z=+0.02$ to avoid overlapping.
Finally, we restrict our comparison to counts in cells of $R=8$ \hmpc since this is the minimum cell size 
considered by K11 and the only one common to the three analyses.

The values  $\hat{b}$  of zCOSMOS galaxies (bottom panel) are in agreement with those of VIPERS galaxies. These values increase with redshift in both cases.
{ This} trend is more evident in the zCOSMOS case, while for VIPERS the evolution is detected only with a significance of 
$\sim 1$ $\sigma$ only.
{ { Our results} do not match those of M05 at $z=0.8,$ where the two samples overlap. The { significance of the} discrepancy, however,
is about 1 $\sigma$. }
{A similar} mismatch was observed between VVDS-Deep zCOSMOS and interpreted by K11 in terms of different clustering {amplitude}  in the two datasets \citep{mccracken07,meneux09,kovac011}. {Indeed, } 
 zCOSMOS  is characterised by 
prominent structures and large spatial coherence as opposed to the VVDS Deep field. This difference was
interpreted as a  manifestation of cosmic variance.
The VIPERS survey was designed to reduce the impact of cosmic variance and solve these types of controversies.  
In this specific case, the agreement between VIPERS and zCOSMOS galaxies  suggests that the 
bias of the latter is closer to the cosmic mean than {that of the} VVDS-Deep field.


The comparison among the non-linear bias parameters of the three
 galaxy samples (upper panel of Fig.~\ref{fig:bias_compf}) corroborates this conclusion. 
 The values of $\tilde{b}/\hat{b}$ for zCOSMOS and VIPERS galaxies agree with each other
 and significantly deviates from unity.
 Thanks to the smaller error bars {in VIPERS} these deviations are now detected
 with higher statistical significance.
 Deviations from non-linear bias in the VVDS-Deep are larger than in VIPERS but the statistical significance 
 for this mismatch is { just about} 1 $\sigma$.
 

\begin{figure}
\includegraphics[width=0.54\textwidth]{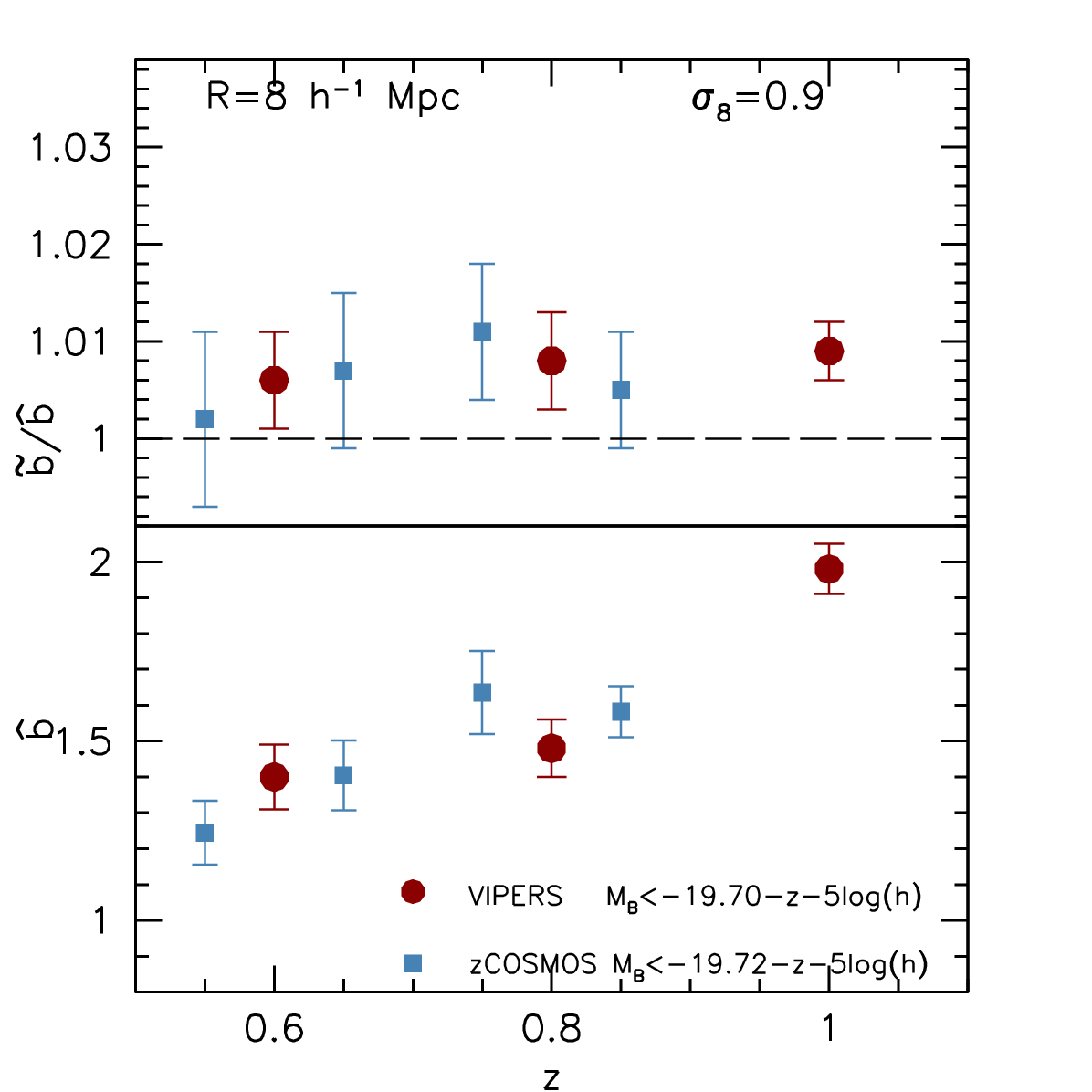}
\caption{Same as figure~\ref{fig:bias_compf}, but referring to brighter
VIPERS (M$_{\rm B}<-19.7-z-5\log(h)$, red dots)
and zCOSMOS galaxies  (M$_{\rm B}<-19.72-z-5\log(h)$, blue squares). 
}
\label{fig:bias_compb}
\end{figure}

Figure~\ref{fig:bias_compb} is analogous to Figure~\ref{fig:bias_compf}. It shows the 
values of $\hat{b}$ and  $\tilde{b}/\hat{b}$ for galaxy subsamples extracted from
VIPERS (red dots) and zCOSMOS (blue squares) using magnitude cuts brighter than before:
 M$_{\rm B}=-19.7-z-5\log(h)$ for VIPERS  
and M$_{\rm B}=-19.72-z-5\log(h)$ for zCOSMOS. 
Our results confirm those obtained with the fainter samples; the values of  $\hat{b}$ and  $\tilde{b}/\hat{b}$
for VIPERS galaxies agree with those of zCOSMOS galaxies in the redshift range in which
the two analyses overlap. Non-linearity is detected at more than 1-$\sigma$ in the  VIPERS sample alone.
No comparison was made with the VVDS-Deep sample in this case
since none of the subsamples analysed by M05 match these luminosity cuts.


\subsection{Linear bias from galaxy clustering}
\label{sec:blincomparison}

In Section~\ref{sec:results} we saw that the bias of VIPERS galaxies 
deviates {from} linearity at all redshifts and on all scales explored.
{  The amount of non-linearity, quantified by the parameter $\tilde{b}/\hat{b}$, 
is rather small, of the order of a few per cent.  { This means that  $\hat{b}$ 
is reasonably similar to $b_{\rm LIN}$ and, therefore, can be compared 
with the linear bias parameter computed in other analyses. }}

In the following, {we therefore compare
 the values of $\hat{b}$ computed in this work
with the values of  $b_{\rm LIN}$ obtained from different datasets in the same redshift range but 
using a variety of bias estimators.}
Galaxy bias at these redshifts has been estimated from both galaxy clustering and weak lensing.
The latter probe, however, has either focused on bright objects used to trace baryonic acoustic oscillations
\citep{comparat13} or to explore bias dependence on the stellar mass \citep{jullo12}.
Therefore, we focus on the values of  $b_{\rm LIN}$ obtained from 
 galaxy clustering in other datasets available in the literature.

\begin{figure}
\includegraphics[width=0.56\textwidth]{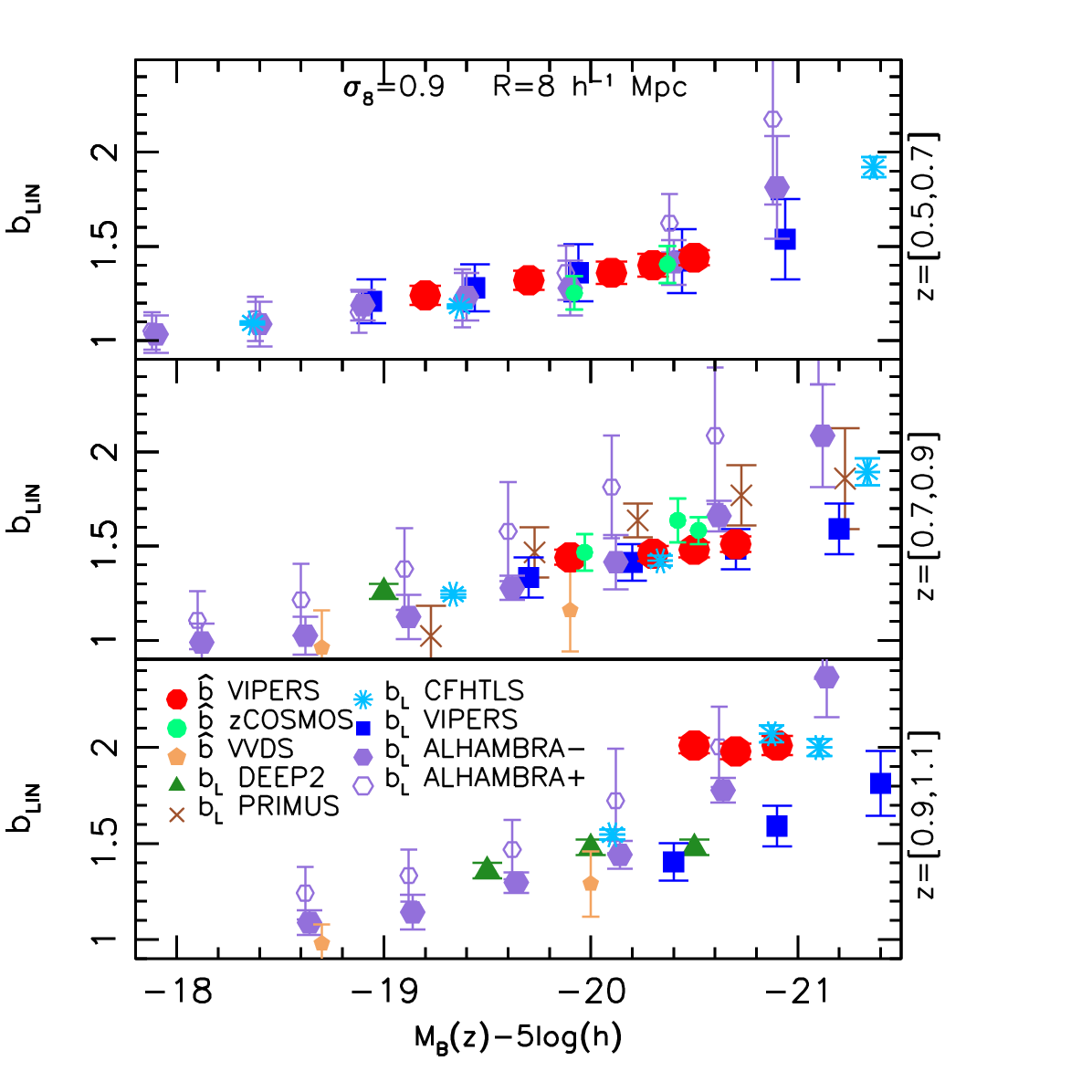}
\caption{Comparison between the bias parameters $\hat{b}$ 
and $b_{LIN}$ obtained from galaxy  counts and galaxy clustering, respectively.
Large red circles: $\hat{b}$ of VIPERS galaxies. Blue squares: $b_{LIN}$ of VIPERS  from 
\cite{marulli13}.  Green triangles: $b_{LIN}$ for DEEP2 galaxies  from \cite{coil06}.
Brown crosses: $b_{LIN}$ for PRIMUS galaxies  from \cite{skibba13}.
Light blue asterisks: $b_{LIN}$ for CHFTLS galaxies  from \cite{coupon12}.
Purple hexagons: $b_{LIN}$ for ALHAMBRA galaxies  from \cite{alhambra13}.Small light green dots: $\hat{b}$ for zCOSMOS galaxies from \cite{kovac011}.
Small light brown pentagons: $\hat{b}$ for VVDS-Deep galaxies from \cite{marinoni05}.
Values of $\hat{b}$ were measured on 
a scale $R=8$ \hmpc.
 }
\label{fig:bias_complin}
\end{figure}

\begin{figure}
\includegraphics[width=0.56\textwidth]{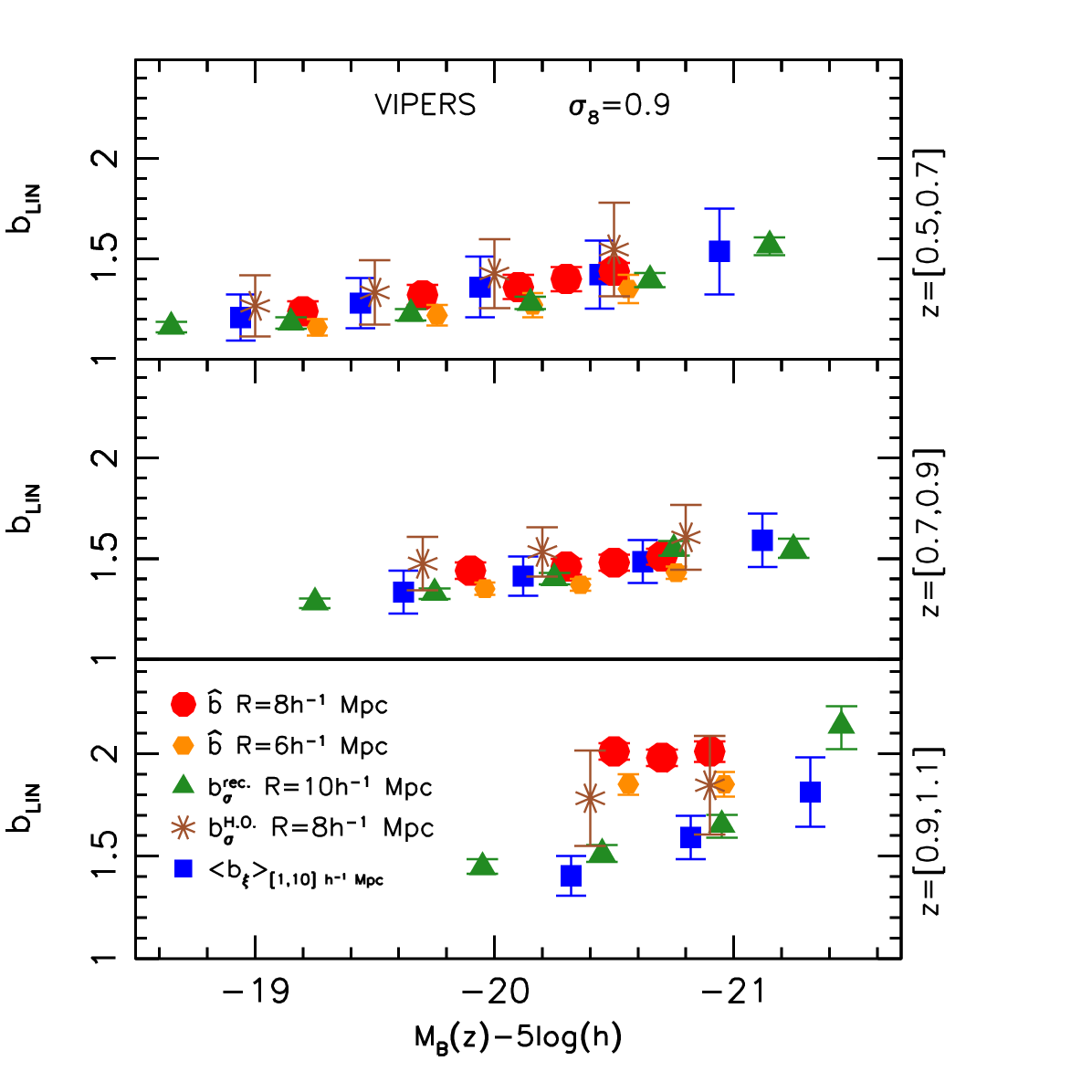}
\caption{{
Comparison between the bias parameter $\hat{b}$ obtained from our analysis
on a scale of $R=8$ \hmpc (large red circles), $ R=6$ \hmpc (small orange hexagons) 
 and  the linear bias parameters  of VIPERS galaxies, $ b_{LIN}$, obtained by 
 \cite{marulli13} (blue squares), \cite{cappi} (brown asterisks), and \cite{granett} (green triangles).}}
\label{fig:bias_vip}
\end{figure}

The results of these comparisons are shown in Figure~\ref{fig:bias_complin}
{ in which we plot the most recent estimates of  both $\hat{b}$ and  $b_{\rm LIN}$
in the three redshift bins as a function of the magnitude cut.
We consider the reference scale of $R=8$ \hmpc since this is the size of the cells 
used to measure $\hat{b}$ in VIPERS (large red circles), VVDS-Deep (small orange pentagons),
 and zCOSMOS (small, light green circles), as shown in the previous section.
 Magnitudes on the X-axis are specified in the B-band {since this band}
 is  used in most of the considered samples  
 with the exception of the PRIMUS and  CHFTLS-wide. For these two { latter} cases,
we consider the $g$-band magnitude and transform it into B-band according to the 
$g-$B versus $z$ relation measured from the VIPERS catalogue.
Finally, all results were normalised to $\sigma_8=0.9$.


The large red circles {represent} the $\hat{b}$ values obtained from VIPERS,  for the  systematic errors listed in Table~\ref{tab:errors}. Therefore { these values} are slightly different from 
those shown in Figures~\ref{fig:bias_compf} and \ref{fig:bias_compb}.

The light blue asterisks represent the $b_{\rm LIN}$ values obtained from the
 Wide part of the Canada-France-Hawaii Legacy Survey (CFHTLS) \citep{coupon12}.
In this case, the bias values were computed from   $\sim 3\times 10^6$ 
galaxies in the redshift interval $z=[0.2,1.2]$ by fitting a Halo Occupation Distribution  model to the measured angular correlation function.
These $b_{\rm LIN}$ values were obtained by integrating the halo bias over the halo mass function. { Therefore they are } integral quantities 
{ much} like  $\hat{b}$, which is computed by integrating over the mass PDF (Eq. \ref{eq:2ndorder}).
 The $b_{\rm LIN}$ values of  CFHTLS galaxies agree well with our results in all redshift bins, including $z=[0.9,1.1]$.

The brown crosses in the middle panel of Figure~\ref{fig:bias_complin} are from \cite{skibba13} and show the bias of
galaxies in the PRIMUS catalogue.
The PRIMUS  \citep{Coil11} galaxy survey is carried out using a low-resolution spectrograph  and complete down to $i < 23$.
This dataset, which  covers five independent fields (including the COSMOS field), spans the redshift range  $z=[0.2,1.0]$.
Here we focus on the interval $z=[0.5,1.0]$ and plot the corresponding bias values in the middle panel.
{ The} bias was estimated from the projected galaxy 2-point correlation function, $w_{p,g}(r)$,
as  $b_{\rm LIN}(r)=\sqrt{w_{p,g}(r)/w_{p,m}(r)}$, where the projected 2-point correlation function of the matter, $w_{p,m}(r)$, 
was modelled assuming a flat $\Lambda$CDM model with $\sigma_8=0.8$.
The bias of PRIMUS galaxies is 
systematically larger than that of VIPERS. However, the significance of the mismatch is below  {\it 1-$\sigma$}.

The purple hexagons in the plot show  $b_{\rm LIN}$ of ALHAMBRA galaxies \citep{alhambra13}.
The photometric redshift survey ALHAMBRA covers seven independent fields, including DEEP2 and COSMOS.
Photometric redshifts are accurate enough to measure the projected galaxy correlation function
at different redshifts and, from this, to estimate the bias.
In Figure~\ref{fig:bias_complin} 
we show the $b_{\rm LIN}$ values estimated in three redshift bins: 
 $[0.35,0.65]$ (top panel), $[0.55,0.85]$ (middle panel), and $[0.75,1.05]$ (bottom panel).
We did not consider  the interval $z=[0.95,1.25]$ { since it is largely beyond the VIPERS range}.
We show two sets of points. Small open hexagons represent the values of $b_{\rm LIN}$
obtained from the clustering of galaxies in all seven fields (labelled ALHAMBRA+ in the plot).
Filled hexagons (labelled ALHAMBRA-) illustrate the effect of removing 
two "outlier" fields, COSMOS and ELAIS-N1, which are characterised by a high degree of  clustering.
The bias of galaxies in ALHAMBRA- agrees with that of VIPERS for $z<0.9$. In the last redshift bin,
for M$_{\rm B}(z=1)<-20.56-5\log(h)$ { the bias of ALHAMBRA- is $\sim 1.5$-$\sigma$ below that of VIPERS.
However, the discrepancy disappears when one considers ALHAMBRA+ and seems to reappear, with a reverse sign, 
at higher luminosities.}

The green triangles show the  $b_{\rm LIN}$ values obtained from 
the projected galaxy 2-point correlation function of
galaxies brighter than M$_B-5\log(h) =-20.5-5\log(h)$
 { at $z=[0.9,1.1]$
 in} the DEEP2 survey \citep{coil06}.
{ In the brightest magnitude bin, where the three samples overlap,
we find that the bias of DEEP2  galaxies is significantly smaller than that of VIPERS and ALHAMBRA
objects}.

{ To summarise,} we find a good agreement {between} the value of $\hat{b}$ measured in our work 
and those of $b_{\rm LIN}$ estimated in a number of surveys in the range  $z=[0.5,0.9]$.
{ In particular,} our measurements agree with those of K11  (small, light green circles)
who used the same technique to estimate $\hat{b}$.

In the outermost redshift shell {not all the bias values measured in different surveys
agree with each other. The value of }
 $b_{\rm LIN}$ for DEEP2 and, to a lesser extent, for
ALHAMBRA- galaxies, are smaller than $\hat{b}$ from VIPERS.
This mismatch may indicate either a genuine difference in the clustering properties of the 
different samples or deviations from linear bias { highlighted by the different bias estimators.}

To {quantify}  the impact of non-linear bias, we compare our $\hat{b}$ values
with {the corresponding $b_{\rm LIN}$ estimated from the very same VIPERS subsamples} considered here.
{ Figure~\ref{fig:bias_vip} compares $\hat{b}$ from VIPERS (red filled symbols) with $b_{\rm LIN}$ from \cite{marulli13}
(blue filled squares, also shown, for reference in Figure~\ref{fig:bias_complin}). 
The two estimates agree at all redshifts but  the last redshift bin
where the bias of \cite{marulli13} matches that of DEEP2 galaxies and, consequently, is significantly below our  $\hat{b}$ value.

Like most of the other measurements, 
\cite{marulli13}  estimated $b_{\rm LIN}$ from the projected 2-point correlation function. More precisely, they averaged
the correlation signal over the interval $r=[1,10]$ \hmpc. In the presence of a scaled dependent bias, a manifestation of which is 
a  $\tilde{b}/\hat{b}$ ratio different from unity, it is not obvious which effective scale of the bias is estimated by \cite{marulli13}.
In our comparison we implicitly assumed that this scale is the same as the cell size, i.e. 8  \hmpc. 
In fact, a small scale seems more appropriate, especially if one accounts for the fact that errors in the projected 
correlation function increases with the pair separation. For this reason, 
we also show $\hat{b}$ measured in cells of $R= 6 $ \hmpc (orange hexagons). 
In this case,  the significance of the mismatch is significantly reduced. 
Decreasing the scale to $R= 4 $ \hmpc (not shown) would bring the two values into agreement
at the price, however, of creating a mismatch at lower redshifts.

Focusing on the VIPERS sample, a  more homogeneous comparison can be performed { considering} the $b_{\rm LIN}$ value 
obtained by \cite{cappi} from}  counts in cells of $R=8 $ \hmpc   (brown asterisks, in the plots).
In this case the results agree with ours within the (rather large) error bars.

{ In the figure we} also show the VIPERS linear bias estimated by \cite{granett} (green triangles) from a Bayesian reconstruction 
of a Wiener filtered, adaptively smoothed galaxy density field. 
The {result agrees with that of \cite{marulli13}. However, as in that case, it is  difficult to associate an effective scale 
to the filtering procedure and perform a homogeneous comparison with our estimate.

Therefore, all the bias estimates of the VIPERS galaxies agree with each other at  $z<0.9$, a sign that galaxy bias is 
largely independent of scales.  At higher redshifts we observe some discrepancies among the 
various estimates whose significance, however, is difficult to assess since the different estimates 
are sensitive to different scales. 
It is safe to conclude that the scale-dependence bias of VIPERS galaxies is more pronounced at  high redshifts,
as confirmed by the results presented in Section~\ref{sec:scale}, and that this can account 
for most of the discrepancies seen in Fig.~\ref{fig:bias_vip}. }

An additional, though minor, {source of discrepancy} is incompleteness. At $z\sim 1$ the 90 \% completeness limit
in VIPERS is  M$_B-5\log(h) \sim -21.0$ for  red galaxies and about half a magnitude fainter for the blue galaxies.
Since red galaxies are more biased than the blue galaxies, selecting objects at this luminosity cut
underestimates the bias of the {composite VIPERS sample}. The amplitude of the effect depends on the luminosity cut.
We conclude that deviations from linear bias cannot be ignored at high redshifts and that
using $b_{\rm LIN}$ as a proxy for galaxy bias leads to significant systematic errors.

}


\section{Discussion and conclusions}
\label{sec:conclusions}

The importance of characterising galaxy bias at intermediate redshifts 
 stems from the need to infer the properties of the distribution of the mass from
that of the galaxies. { This will be especially important in } future redshift surveys aimed at an accurate estimate of 
the cosmological parameters.
This  has {prompted} several  { efforts to estimate galaxy bias}
at $z>0.5$ { exploiting} weak lensing, galaxy clustering, and galaxy counts.
Most of these works assume that galaxy bias is linear and deterministic and 
provide an estimate for the linear bias parameter.
In this work we questioned this assumption and searched for possible deviations from linear bias.
This issue has already been investigated by M05 and K11 (using counts in cells { and significantly smaller samples}) and by
\cite{simon07} and \cite{jullo12} {  with conflicting evidence, as discussed in the introduction}.

Our work builds upon these results improving the original strategy of M05 and K11 
in several aspects.
First of all, it is based on a  new dataset of $\sim 50,000$ galaxies distributed over a much larger volume
than its predecessors. This significantly 
reduces the impact of cosmic variance that in previous studies dominated the error budget.
Secondly, we use a new technique to infer the mean biasing function from counts in cells that, under
the hypothesis of local Poisson sampling,  accounts and automatically corrects for shot noise.
This improvement greatly increases our ability to recover the biasing function since Poisson noise is
the main source of stochasticity in the bias relation.
Thirdly, owing to the size of the sample, we are able to explore the bias dependence on 
magnitude, redshift, and scale with unprecedented accuracy. We postpone the investigation  
of additional dependences on galaxy colour and stellar mass
to a future analysis to be performed with the final VIPERS sample and new mock galaxy catalogues 
designed to mimic these galaxy properties.

The main results of our study are:

The overall qualitative behaviour of  the mean biasing function of VIPERS galaxies is 
similar to that of zCOSMOS and VVDS-Deep galaxies as well as to that of the 
synthetic VIPERS galaxies in the mock catalogues  that we used to estimate errors. The shape of the mean biasing function is close to linear in regions above the mean density. 
{ It deviates from linear bias at  $\delta<0$. More specifically, above the threshold $\delta_{\rm TH}$ at which $\delta_g=-0.9,$ 
the bias function is significantly steeper than its mean slope $\hat{b}$ on scales smaller than 8 \hmpc. For over-densities { below} $\delta_{\rm TH}$ the mean biasing function
features a tail that cannot be accounted for by linear biasing. 
The over-density threshold $\delta_{\rm TH}$ has been interpreted as a typical 
density scale below which very few galaxies form. In our analysis, we find that this threshold
 increases with the redshift and with the luminosity cut-of,f 
so that at moderate redshifts low-density regions are typically populated by faint galaxies.}

The biasing function shows small but significant deviations { from {linearity}}
at all redshifts, scales, and magnitude 
intervals that we explored. The parameter $\tilde{b}/\hat{b}$ that we use to 
quantify non-linearity neither seems to evolve with the redshift nor to depend on the 
luminosity. A scale dependence is observed at low redshifts  below  6 \hmpc
{ with only $\sim 1 \, \sigma$ significance}.
  
We confirm that galaxy bias depends on luminosity.
The mean slope $\hat{b}$ of the biasing function, a { good} proxy to linear  bias
given the small degree of non-linearity,
increases with the luminosity threshold used to select the 
galaxy sample. The effect is
significant for  $z=[0.5,0.7]$, probably thanks to the 
large magnitude leverage here compared to the bin   $z=[0.9,1.1]$, {in} which we can
 probe a { much} smaller magnitude range $\Delta M_{\rm B}=0.4$.
The value of $\delta_{\rm TH}$ also increases with the magnitude, suggesting that 
the efficiency of galaxy formation decreases with the luminosity of the object.

We confirm that galaxy bias increases with redshifts as predicted
by most bias models and verified in other datasets. In our case we
find evidence for a rapid evolution beyond $z=0.9$. 
This result is highly significant and robust 
since it depends neither on the scale nor on the luminosity of the objects.
The  statistical significance of this result depends on the reliability of our error analysis,
which is based on a mock galaxy catalogue designed to match the correlation properties 
of VIPERS galaxies. { { We verified that mock catalogues are} very realistic in the sense that 
{their} biasing function  matches that of  real objects remarkably well.}
In this analysis, we modelled 
all known sources of systematic errors, including the magnitude dependence of the 
spectroscopic sampling rate that was not originally included in the mock catalogues.
We find no evidence for systematic errors that might 
{ mimic a spurious evolution in the bias moments $\hat{b}$ and $\tilde{b}$}.

The value of  $\hat{b}$ increases with the scale from 4 to 8 \hmpc.We interpret this  in the framework of the halo model as the transition between 
the one-halo and two-halo contribution to galaxy bias. 
The same trend is seen in our mock catalogues, in which objects were extracted 
assuming the HOD model and 
in previous analyses performed at lower redshifts using both galaxy clustering  (e.g. \cite{zehavi05})
and weak lensing (e.g. \cite{hoekstra02}, \cite{simon07}).
This is the first time that this effect is detected
at high redshifts with counts in cells statistics.

We compared our results with those of M05 and K11. These
authors performed an analysis similar to that presented
here in a similar range of redshifts. We limited the comparison
on a scale of 8 \hmpc, {which is} common to  the three analyses.
We find that the values of $\hat{b}$ of VIPERS and zCOSMOS galaxies agree 
within the errors. M05 find a smaller degree of biasing but the 
difference is of the order of  $1-\sigma$. 
We conclude that the claimed discrepancy between K11 and M05 results 
is a manifestation of cosmic variance.

Deviations from linear biasing were also detected by M05 and K11, 
although with a lower significance than in our case. Our results agree with those of K11.
In M05 the degree of non-linearity is slightly
larger than in our case but the discrepancy is barely larger than 
1-$\sigma$. 
The bias non-linearity is sometimes expressed in terms of the  parameter $b_2$
of the second-order Tailor expansion of $\delta$ \citep{fg93}. 
\cite{cappi} analysed the same VIPERS dataset  using higher order statistics, which is a procedure
that is less sensitive to non-linear bias than ours. They detected a deviation from linear 
bias at  $z  \le 0.9$ with a significance of $\sim 1$-$ \sigma$. Their  $b_2$ 
value turned out to be negative, in agreement with M05 and
\cite{marinoni08} and, qualitatively,  with our results too.

measured from the clustering of galaxies in recent galaxy redshift surveys 
(DEEP2,  PRIMUS,  CHFTLS-wide, and ALHAMBRA).
This comparison is qualitative since it assumes that bias is linear, while our analysis has detected a 
small, but significant, degree of non-linearity in the bias of VIPERS galaxies.
The comparison is generally successful at $z<0.9,$ where we find a very good agreement with 
all existing results. In this redshift range our results provide additional evidence in favour of
a  luminosity-dependent bias and of a weak evolution. 
At $z>0.9$, where the spread among current results is large,
our results favour the case of a significant bias evolution, in agreement with the  CHFTLS-wide and
ALHAMBRA analyses.


Our  results confirm the importance of  going beyond the simplistic linear biasing hypothesis.
Galaxy bias is a complicated phenomenon. It can be non-deterministic, non-local, and non-linear.
In this work we focused on deviations from linearity under the assumption that stochasticity is dominated by (and, 
consequently, accounted for) Poisson noise and that non-local effects are smoothed out within the volume of our cells.
While the validity and the impact of these assumptions can (and will) need to be tested,
our results show that the application of an improved statistical tool to the new VIPERS dataset is already able to detect 
deviations from linear bias  with $5-10$ \% accuracy.


\section*{Acknowledgments}

We acknowledge the crucial contribution of the ESO staff for the management of service observations. In particular, we are deeply grateful to M. Hilker for his constant help and support of this programme. Italian participation in VIPERS is funded by INAF through PRIN 2008 and 2010 programmes. LG and BRG acknowledge support of the European Research Council through the Darklight ERC Advanced Research Grant (\# 291521). OLF acknowledges support of the European Research Council through the EARLY ERC Advanced Research Grant (\# 268107). Polish participants are supported by the Polish Ministry of Science (grant N N203 51 29 38), the Polish-Swiss Astro Project (co-financed by a grant from Switzerland, through the Swiss Contribution to the enlarged European Union), the European Associated Laboratory Astrophysics Poland-France HECOLS, and a Japan Society for the Promotion of Science (JSPS) Postdoctoral Fellowship for Foreign Researchers (P11802). GDL acknowledges financial support from the European Research Council under the European Community's Seventh Framework Programme (FP7/2007-2013)/ERC grant agreement n. 202781. WJP and RT acknowledge financial support from the European Research Council under the European Community's Seventh Framework Programme (FP7/2007-2013)/ERC grant agreement n. 202686. WJP is also grateful for support from the UK Science and Technology Facilities Council through the grant ST/I001204/1. EB, FM, and LM acknowledge the support from grants ASI-INAF I/023/12/0 and PRIN MIUR 2010-2011. LM also acknowledges financial support from PRIN INAF 2012. YM acknowledges support from CNRS/INSU (Institut National des Sciences de l'Univers) and the Programme National Galaxies et Cosmologie (PNCG). CM is grateful for support from specific project funding of the { Institut Universitaire de France} and the LABEX OCEVU. MV is supported by FP7-ERC grant ``cosmoIGM'' and by PD-INFN INDARK.
CDP wishes to thank Pangea Formazione S.r.l. for supporting her in the final stages of this work.




\end{document}